\documentclass[fleqn,usenatbib]{mnras}

\usepackage[T1]{fontenc}

\DeclareRobustCommand{\VAN}[3]{#2}
\let\VANthebibliography\thebibliography
\def\thebibliography{\DeclareRobustCommand{\VAN}[3]{##3}\VANthebibliography}

\usepackage{graphicx}	
\usepackage{amsmath}	
\usepackage{amssymb}	
\usepackage{newtxtext,newtxmath}

\title[Ram-pressure stripping in A1367]{MUSE sneaks a peek at extreme ram-pressure stripping events - V. Towards a complete view of the galaxy cluster A1367}

\author[A.Pedrini et al.]{
Alex Pedrini$^{1}$\thanks{E-mail: alex.pedrini@campus.unimib.it},
Matteo Fossati$^{1,2}$\thanks{E-mail: matteo.fossati@unimib.it},
Giuseppe Gavazzi$^{1}$,
Michele Fumagalli$^{1,3}$,
Alessandro Boselli$^{4}$,\and
Guido Consolandi$^{1}$,
Ming Sun$^{5}$,
Masafumi Yagi$^{6}$,
Michitoshi Yoshida$^{7}$
\\
$^{1}$Dipartimento di Fisica G. Occhialini, Universit\`a degli Studi di Milano-Bicocca, Piazza della Scienza 3, 20126 Milano, Italy \\
$^{2}$INAF-Osservatorio Astronomico di Brera, via Brera 28, I-20121 Milano, Italy \\
$^{3}$INAF - Osservatorio Astronomico di Trieste, via G. B. Tiepolo 11, 34143 Trieste, Italy \\
$^{4}$Aix Marseille Univ, CNRS, CNES, LAM, Marseille, France \\
$^{5}$Department of Physics and Astronomy, University of Alabama in Huntsville, Huntsville, AL 35899, USA \\
$^{6}$National Astronomical Observatory of Japan, 2-21-1, Osawa, Mitaka, Tokyo 181-8588, Japan \\
$^{7}$Subaru Telescope, National Astronomical Observatory of Japan, 650 North A'ohoku Place, Hilo, HI 96720, USA }

\date{Accepted 2022 February 4. Received 2022 February 4; in original form 2021 November 15}

\pubyear{2021}

\begin{document}
\label{firstpage}
\pagerange{\pageref{firstpage}--\pageref{lastpage}}
\maketitle

\begin{abstract}
We present an analysis of the kinematics and ionization conditions in a sample composed of seven star-forming galaxies undergoing ram-pressure stripping in the A1367 cluster, and the galaxy ESO137-001 in the Norma cluster. MUSE observations of two new galaxies in this sample, CGCG097-073 and CGCG097-079, are also presented.  This sample is characterized by homogeneous integral field spectroscopy with MUSE and by a consistent selection based on the presence of ionised gas tails. The ratio [OI]/H$\alpha$ is consistently elevated in the tails of these objects compared to what observed in unperturbed galaxy disks, an ubiquitous feature which we attribute to  shocks or turbulent phenomena in the stripped gas. Compact star-forming regions are observed in only $\approx 50 \%$ of the tails, implying that specific (currently unknown) conditions are needed to trigger star formation inside the stripped gas. Focusing on the interface regions between the interstellar and intracluster medium, we observe different line ratios that we associate to different stages of the stripping process, with galaxies at an early stage of perturbation showing more prominent signatures of elevated star formation.
Our analysis thus demonstrates the power of a well selected and homogeneous sample to infer general properties arising from ram-pressure stripping inside local clusters.
\end{abstract}

\begin{keywords}
ISM: evolution – galaxies: clusters: intracluster medium – galaxies: evolution – galaxies: clusters: individual: A1367 – galaxies: interactions - galaxies: star formation
\end{keywords}



\section{Introduction}
Within the general framework of galaxy evolution, the environment is known to play a key role in shaping the galaxy life cycle. In particular, the environment can actively remove the  gas reservoir from a galaxy or lead to its consumption, ultimately quenching the star formation activity and turning blue objects into red and dead ones. Since the seminal work of \citet{Dressler80}, and later with the advent of large multi wavelength surveys (e.g. the Sloan Digital Sky Survey, SDSS \citealt{York00}) it has become apparent that the properties of galaxies in groups and clusters are substantially different compared to their counterparts in the field \citep{Dressler97, Baldry06, Fumagalli09, Peng10, Wetzel13}. Specifically, massive halos host a large fraction of red ellipticals and less blue spirals compared to the field \citep{Whitmore93}.
Differently from isolated galaxies, where only internal processes determine their evolution, galaxies in dense environments are potentially subject to a large number of physical processes. These can be divided in two main subclasses: gravitational interactions between galaxies or with the host halo potential such as tidal interactions or galaxy harassment \citep{Merritt83,Byrd90,Moore96}, and hydrodynamic interactions acting between the interstellar medium (ISM) of a galaxy moving at high velocity through the hot and dense plasma of the host halo (the intra cluster medium, ICM). Among the hydrodynamic processes there are ram-pressure stripping \citep[RPS;][]{Gunn72}, viscous stripping \citep{Nulsen82}, thermal evaporation \citep{Cowie77} and starvation \citep{Larson80}.

In this picture, if the gas reservoir is being removed from the galaxy as opposed to being consumed by star formation, gas can be observed in various phases usually distributed in ``cometary tails'' downstream of the parent galaxy. In the last decades, several observations have unveiled these tails at different wavelengths, such as in X-ray \citep[e.g.][]{Wang04, Sun10, Zhang13, Schellenberger15}, in the H$\alpha$ emission line \citep[e.g.][]{Kenney95, Gavazzi95, Conselice01, Yoshida02, Sun07, Kenney14, Boselli16, Yagi17},  in HI \citep[e.g.][]{Oosterloo05, Hota07, Chung07, Scott13}, in radio continuum \citep[e.g.][]{Gavazzi78, Miley80, Dickey84, Hummel91, Roberts21}, and from the analysis of molecular lines \citep[e.g.][]{Boselli94, Salome06, Scott13, Jachym14, Scott15}.

Despite this multitude of observations, we have yet to obtain a full understanding of what physical processes are the most relevant in dense environments, what is their interplay and how they transform galaxies of different stellar mass and morphology. Recent observations of entire clusters \citep{Yagi10, Yagi17, Boselli16, Boselli18} 
suggest that ram-pressure stripping is almost ubiquitous in local massive clusters and has high efficiency in quenching galaxies removing their ISM on time scales comparable with (or shorter than) their cluster crossing time \citep{Boselli21REW}. The sample from the GASP survey \citep{Poggianti17} also revealed a significant fraction of galaxies with RPS tails among cluster galaxies.
Furthermore, this mechanism is sometimes able to temporarily enhance the star formation in the interface region of galaxies as indicated from observations or simulations \citep[e.g.][]{Boselli94, Bekki03, Steinhauser12, Scott15, Vulcani18, Lee20, TroncosoIribarren20, Boselli21}. As a result, the study of galaxies suffering RPS has become crucial in order to understand the general evolution and physical properties of cluster galaxies.

While extended RPS tails can be observed at several wavelengths and gas phases, an ideal tracer for the stripping phenomenon is the ionised gas around galaxies which can be observed in the optical combining the high spatial resolution and sensitivity of ground-based telescopes by means of observations of the H$\alpha$ emission line \citep{Fossati12, Fumagalli14, Gavazzi17, Boselli18b}. Significant progress in the study of this process has been enabled by recent observations of entire clusters at great depth with wide-field cameras \citep{Fasano06, Yagi10, Yagi17}. Despite their largest discovery power, however, narrow-band imaging observations of the H$\alpha$ line do not provide stringent constraints on the physics of the stripping phenomenon, for which it is critical to have access to the kinematics and the ionization conditions of the gas from spectroscopic observations.  First attempts to collect spectra of these faint and extended filaments have been carried out, e.g., by \citet{Cortese06} and \citet{Yoshida08, Yoshida12} using traditional long-slit spectroscopic observations. These efforts, however, provide modest spatial information and are often limited to the brightest filaments. A revolution in this field has been brought by the deployment of integral field spectroscopic instruments, like the Multi Unit Spectroscopic Explorer (MUSE; \citealt{Bacon10}) at the ESO Very Large Telescope (VLT),  thanks to their ability to map extended sources thus reducing slit losses and reaching the more typical surface brightness levels of RPS tails.

Using MUSE, we have carried out a survey of RPS events in the Norma and A1367 clusters which we have presented in previous papers of this series \citep{Fumagalli14, Fossati16, Consolandi17, Fossati19}. This sample has been selected by narrow-band imaging observations revealing long tails of ionized gas trailing behind cluster galaxies. Specifically, in this sample of galaxies suffering RPS we include ESO137-001, which is falling in the Norma cluster presenting a double tail of diffuse gas, and CGCG097-087 in A1367 which is travelling edge-on through the ICM of the cluster with its smaller companion CGCG097-087N. We also include the Blue Infalling Group (BIG) found near the center of A1367 which is composed of three massive galaxies (two of which suffering from RPS) CGCG097-120, CGCG097-114 and CGCG097-125 and several dwarf systems formed by the tidal interactions in the group. Adding to this sample, in this work we present the analysis of the new MUSE observations of two low-mass galaxies CGCG097-079 and CGCG097-073 undergoing RPS in A1367. These new observations allow us to obtain a statistical view of RPS among star-forming galaxies in this rich cluster, enabling the study of a galaxy sample (as opposed to individual objects) and paving the road to future work to understand the physical properties of the galaxies and the diffuse gas around them in large samples.

The paper is structured as follows. After a presentation of the A1367 galaxy cluster in Section~\ref{sec:cluster}, where we introduce the main features that make this environment particularly rich in perturbed objects, we present novel data for two cluster members, CGCG097-079 and CGCG097-073. We first present the data reduction procedures and the techniques used to derive the stellar continuum from the galaxies and maps of pure emission lines in Section~\ref{sec:dr}. We present the results of an in-depth study of these two new objects in Section~\ref{sec:newgal}. We then combine all these homogeneous MUSE observations into a coherent sample and we discuss our results in the broader context of galaxies suffering RPS in clusters in Sections~\ref{sec:emlines} and \ref{sec:discussion}. Our conclusions can be found in Section~\ref{sec:conclusion}.

Throughout, we adopt a \citet{Chabrier03} stellar initial mass function and we assume a flat $\Lambda$CDM Universe with $\Omega_M = 0.3$, $\Omega_{\Lambda} = 0.7$ and $H_0 = 70$ km s$^{-1}$ Mpc$^{-1}$. With the adopted cosmology, the comoving radial distance of A1367 becomes 92.2 Mpc \citep{Boselli21REW} and 1~arcsec on the sky corresponds to 0.45 kpc.

\section{Abell 1367 and its galaxy population}
\label{sec:cluster}
\begin{figure*}
	\includegraphics[width=0.95\textwidth]{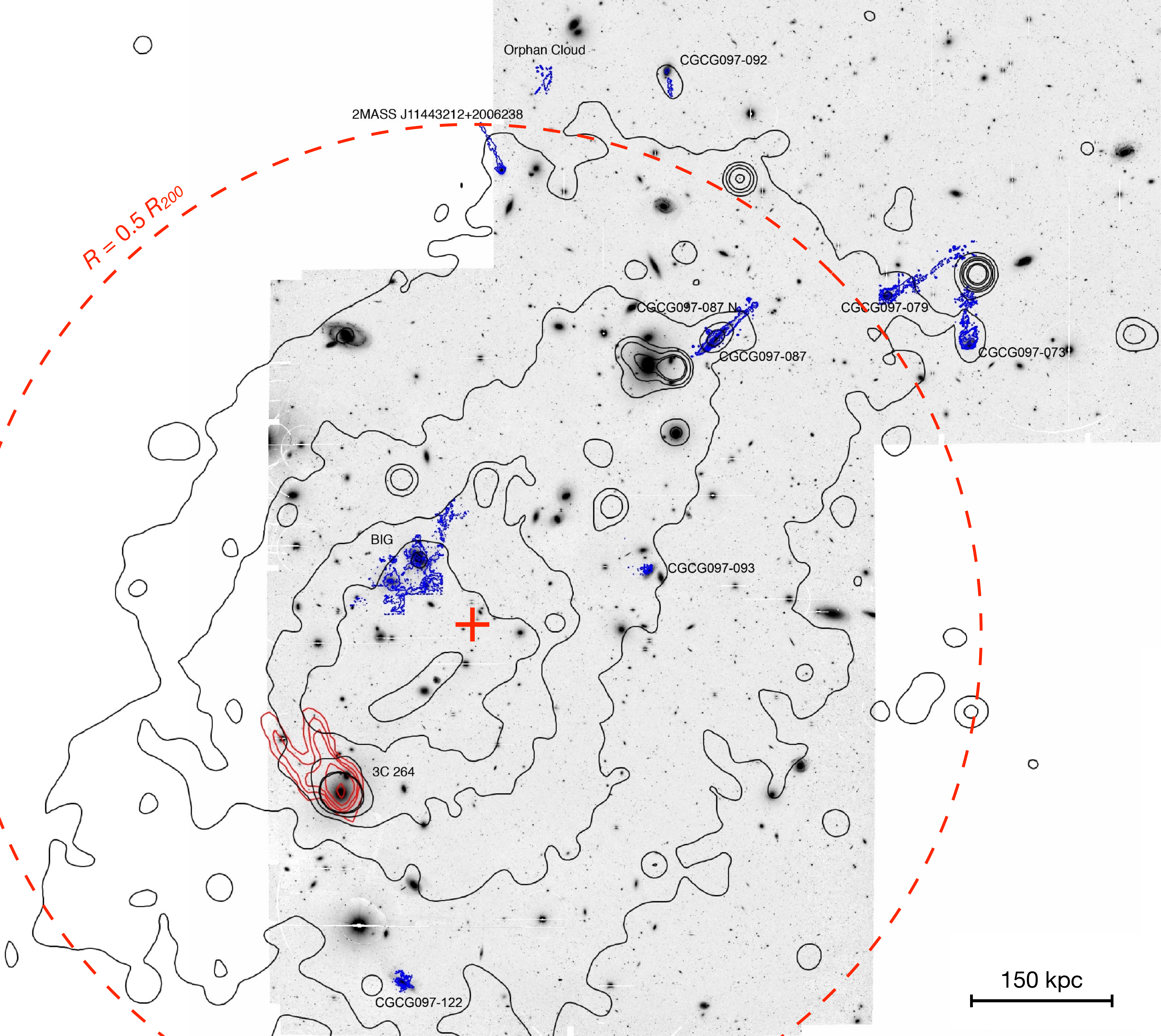}
    \caption{The galaxy cluster Abell1367: in grey scale, the $B$-band Subaru image of the cluster \citep{Yagi17}; in black, X-ray contours from XMM observations of A1367; in blue, the H$\alpha$ flux contours of the extended H$\alpha$ emitters observed with Subaru \citep{Yagi17}, MUSE \citep{Consolandi17, Fossati19} and the 2.12 m telescope of San Pedro Martir \citep{Gavazzi17}; in red, the VLA 21 cm radio continuum contours of the head-tail radio galaxy NGC 3862 \citep{Gavazzi81}. The red cross marks the X-Ray cluster center and the dashed circle shows $R=0.5R_{200}$ from \citet{Boselli21REW}.}
    \label{fig:A1367}
\end{figure*}

The cluster Abell 1367, located at a distance of $92.2 \pm 1.2$ Mpc and with an average recessional velocity of $6484 \pm 81$ ${\rm km~s^{-1}}$ \citep{Cortese04}, is an excellent laboratory to study galaxies currently ongoing ram-pressure stripping. In particular, the presence of an high fraction of spiral galaxies \citep{Butcher84} together with the cluster ongoing merging of two substructures makes this system one of the best  environments to search for tails of ionised gas outside galaxies. In Fig.~\ref{fig:A1367} we present a map of this rich cluster, showing the X-rays distribution (black contours, \citealt{Forman03}) overlaid on the $B-$band Subaru image of the cluster \citep{Yagi17}. Individual galaxies with ionised gas tails (studied in this series of papers and in \citealt{Gavazzi17}) are shown in blue, while a head-tail elliptical radio-galaxy is shown in red near the cluster center. Specifically, near the center of A1367 it is possible to notice the Blue Infalling Group (BIG) and more to the NW we see CGCG097-087 with its companion CGCG097-087N and the pair composed by CGCG097-079 and CGCG097-073. 

The cluster is highly elongated in the NW-SE direction both in the optical distribution of the galaxies \citep{Cortese04} as well as in the X-rays \citep{Sun02}. This evidence has been interpreted as due to two merging substructures where the NW one is rich of young star-forming galaxies \citep{Donnelly98, Cortese04}. A shock front \citep{Ge19} arising from the merger is roughly at the position where most of the RPS galaxies are located in projection. It is therefore possible that the significant number of galaxies undergoing RPS in the NW region of the cluster is also due to a higher density in the hot gas as the shock propagates outwards, enhancing the local ram-pressure above what typical for comparable distances from the cluster center. Finally, observations  in 120-168 MHz radio continuum \citep{Gavazzi95,Roberts21} reveal that four of the galaxies we studied (namely CGCG097-073, CGCG097-079, CGCG097-114 and CGCG097-087) have also clear radio continuum extended tails, further reinforcing the interpretation of RPS being at work in this sample. For completeness, although it is not the subject of this work, we mention that near the cluster center lies NGC3862 (3C264), a well-known strong head-tail radio galaxy (red contours in Fig.~\ref{fig:A1367}, \citealt{Gavazzi81}) showing that some form of RPS is taking place also for more massive and passive galaxies.

As a result of the ongoing merger of the two sub-structures, the cluster is not yet virialized. Estimates of the halo mass are therefore more uncertain compared to more relaxed clusters, but there is a good agreement between different works \citep{Girardi98, Rines03, Cortese04, Boselli21REW} and are in the halo mass range of $M_{\rm halo} \approx 3-7 \times 10^{14} \rm M_\odot$.
Using the sample in \citet{Gavazzi10} and later updated in Gavazzi et al. in prep. based on a catalogue drawn from SDSS, we find 174 galaxies with $M_* > 10^9~\rm{M_\odot}$ within a $R_{\rm vir}$ of A1367 (R$_{\rm vir} = \rm R_{200} \approx 1.4~\rm Mpc$;  \citealt{Cortese04, Boselli21REW}). Of these, 47 (27\%) are blue and star forming according to a $NUV-i$ color selection (corrected for internal extinction). \citet{Boselli&Gavazzi14} and \citet{Boselli21REW} show that 11 star forming galaxies (23\%) show signs of RPS (as seen from either ionized gas or radio continuum observations).

\subsection{The RPS galaxy population}

With the advent of MUSE, in the last 5 years we have observed  $7/11$ ($63 \%$) of the galaxies suffering RPS in A1367 with deep integral field spectroscopy. While the RPS signatures come from observations at different wavelengths, all the objects we followed-up with MUSE have ionized gas tails longer than twice the size of the stellar disk from \citet{Yagi17}. These authors observed the cluster up to $R \approx 0.6 \times R_{\rm vir}$ down to a surface brightness limit of $2 \times 10^{-18}~{\rm erg~s^{-1}~cm^{-2}~arcsec^{-2}}$ finding 10 objects. Requiring a long ionized gas tail (longer than twice the disk size) leads to 8 objects, of which we have observed 6, plus CGCG097-120 which is in the BIG field and is a face-on RPS event \citep{Fossati19}.

The consistent selection coupled with the uniform IFU follow-up with MUSE makes our sample highly homogeneous and representative of RPS objects in the A1367 cluster. Our sample includes CGCG097-087 (a.k.a. UGC6697) and CGCG097-087N \citep{Consolandi17} and BIG \citep{Fossati19} as discussed in detail in previous papers of this series. Here, we include also new observations of CGCG097-079 and CGCG097-073, and analyse them in Section~\ref{sec:newgal}. We now briefly overview the properties of this sample, also summarizing the salient details of the MUSE analysis presented in previous papers.

CGCG097-087 is the second brightest spiral/Irr member of A1367, transiting almost edge-on through the ICM towards the cluster center. Its smaller companion (CGCG097-087N) lies at $\sim$ 20 arcsec in the NE direction from CGCG097-087 and recent Subaru observations \citep{Yagi17} were highly suggestive of galaxy-galaxy interactions due to double gaseous tails connecting the two galaxies. CGCC097-087 and CGCG097-087N have a stellar mass of $\log(M_*/\rm M_{\odot}) = 10.13$ and $\log(M_*/\rm M_{\odot}) = 8.78$ respectively, computed from $g-$ and $i-$ band photometry following the colour calibrations by \citet{Zibetti09}. 
The analysis of this galaxy pair using the high sensitivity and spatial resolution of MUSE gave us unique new information about the hydrodynamic processes driving the stripping and the interaction of the two systems, and most importantly about the physics in the tails. \citet{Consolandi17} found that CGCG097-087 shows complex gas kinematics due to the superposition of gas with different velocity components along the line of sight. Specifically, the ionised gas is split in a component still bound to the galaxy and in a stripped component that is consistent with a RPS event seen edge-on. Furthermore, also CGCG097-087N is suffering RPS from the ICM of the cluster, even if kinematics indicators suggest also a past galaxy-galaxy interaction.  

In the more complex scenario of ram-pressure stripping and ``pre-processing'', i.e. the environmental transformation occurring within galaxy groups before they are accreated onto galaxy clusters, we find the Blue Infalling Group, also known as BIG. Studied by \citet{Sakai02, Gavazzi03, Cortese06,Ge21}, the more recent observations with MUSE \citep{Fossati19} confirm that BIG is an exceptional system in order to understand the role of different processes in the group environment.
The BIG is mainly composed of two galaxies CGCG097-125 and CGCG097-114 with mass of $\log(M_*/\rm M_{\odot}) \approx 10.32$, $\log(M_*/\rm M_{\odot}) \approx 9.22$ respectively. A third galaxy CGCG097-120 ($\log(M_*/\rm M_{\odot}) \approx 10.53$) is unlikely to be part of the group due to its offset in recessional velocity compared to the other galaxies and filaments in BIG \citep{Fossati19}. Furthermore, we find lower mass star-forming systems connected by ionised gas filaments, which have originated from strong tides during the merger of the galaxies that now compose CGCG097-125.
Both CGCG097-120 and CGCG097-114 are galaxies suffering RPS. CGCG097-120, despite not being part of the group, gives us one of the best known example of face-on RPS in the cluster potential. During our analysis in Section~\ref{sec:emlines}, we thus exclude CGCG097-125 as its highly perturbed stellar and gaseous kinematics is likely to arise more from a past merger.

CGCG097-073 and CGCG097-079 \citep{Zwicky68} are two blue, late-type galaxies of $\log(M_*/\rm M_{\odot}) \approx 9.27$ and $\log(M_*/\rm M_{\odot}) \approx 9.08$ respectively, which are likely to be infalling for the first time into the cluster environment \citep{Gavazzi01}. Their recessional velocities are $7298\rm~km~s^{-1}$ and $7025\rm~km~s^{-1}$, respectively or $+814\rm~km~s^{-1}$ and $+541\rm~km~s^{-1}$ compared to the average velocity of the cluster. They lie well within the A1367 virial radius, being at a projected distance of $\approx0.5 \times \rm R_{vir}$. CGCG097-073 is face-on and appears as a spiral galaxy while CGCG097-079 is possibly seen edge-on but its morphology appears irregular, with no evident spiral arms. These galaxies were selected from the works in radio continuum of \citet{Gavazzi78} and \citet{Gavazzi87}, who discovered their head-tail radio morphology indicative of an ongoing RPS event.
Since these discovery papers, CGCG097-073 and CGCG097-079 have been extensively studied at all wavelengths, including in optical \citep{Gavazzi95, Yagi17}, H$\alpha$ 
\citep{Kennicutt84, Gavazzi01, Yagi17}, near infrared \citep{Gavazzi96}, CO \citep{Boselli94, Sivanandam14, Scott13, Scott15}, 
HI \citep{Gavazzi06, Scott10}, and radio continuum \citep{Gavazzi95, Scott10}. All works consistently indicate that both galaxies are suffering 
RPS from the ICM of A1367. 
In particular, the H$\alpha$ narrow-band imaging works by \citet{Gavazzi01} and \citet{Yagi17} revealed that the ionised gas of CGCG097-073 
and CGCG097-079 extends in 
two tails of $\approx 70 \rm ~kpc$ and $\approx 100 \rm ~kpc$, respectively. The orientation of the two tails suggests a $\approx$ 60 degree motion between the two galaxies in the plane of the sky.
The two tails are convergent and although there is no clear spatial continuity in the flux of the two tails even in the deep Subaru observations of \citet{Yagi17}, it cannot yet be excluded that the two galaxies may have interacted in the past \citep{Gavazzi01}. We will discuss in Section~\ref{sec:dis7379} if the MUSE observations support this scenario or whether we are observing only a superposition of two galaxies in the dense cluster environment. 

We also include in the following sections data from ESO137-001 \citep{Fumagalli14, Fossati16} which has similar high-quality MUSE data and is falling in the Norma cluster which has a similar halo mass ($M_{\rm halo,Norma} \approx 2-3 \times M_{\rm halo,A1367}$, \citealt{Cortese04, Woudt08, Boselli21REW}). ESO137-001 is suffering extreme RPS showing a double tail which extends for at least $80$ kpc and with a large number of bright compact HII regions, located where the gas is dynamically cool.   

Thus, altogether, we have collected a statistical and homogeneous sample of galaxies suffering RPS in the mass range $\approx 10^{9}-10^{10.5}~\rm M_\odot$. The physical properties of the galaxies in our sample are listed in Table \ref{tab:sample}.

\begin{table*}
\caption{The sample of RPS galaxies used in this work. The table lists: the galaxy name; the name of the parent cluster; the nuclear activity based on optical line ratios ([OIII]/H$\beta$ vs. [NII]/H$\alpha$); the galaxy stellar mass; the projected distance from the cluster center normalized by the cluster $R_{200}$; the radial heliocentric velocity; the HI-deficiency parameter. A full description of the origin of these values is given in Appendix A of \citet{Boselli21REW}. }
{
\begin{tabular}{lcccccrl}
\hline

Name      	& Cluster		    & Nuc. & log($M_{*}/\rm M_\odot)$ &  $R/R_{200}$& c$z$ (km s$^{-1}$)	    &       HI-def	        \\  
\hline
\hline

ESO137-001	& Norma		    &    HII    & 9.81				 & 0.18		& 4461		& 	-	\\
CGCG097-087	& A1367		    &   HII     & 10.13			 & 0.38		& 6727		&	0.00 	\\
CGCG097-087N    & A1367           &   HII     & 8.78          & 0.37      & 7542       &   - \\
CGCG097-073	& A1367		    &   HII     & 9.27			     & 0.57		& 7298		&  -0.09 	\\ 
CGCG097-079	& A1367		    &   HII     & 9.08				 & 0.52		& 7026		&   0.17 		\\
CGCG097-114 & A1367           &   HII     & 9.22               & 0.05      & 8425      &  -0.32     \\
CGCG097-120 & A1367            &   LIN     & 10.53              & 0.08      & 5620      &   0.90     \\
CGCG097-125 & A1367           &   HII     & 10.32              & 0.08      & 8311      &  -0.05      \\
\noalign{\smallskip}
\hline
\end{tabular}

\label{tab:sample}}
\end{table*}

In this paper we want to fully understand what are the different physical properties leading to different observational features in this sample, constraining the role of RPS in rich nearby clusters. We will also discuss to what extent different ionization stages in these galaxies and their tails can be explained as evolutionary sequence due to different phases of the RPS interaction. 

Before proceeding with the analysis of the whole sample, in Section~\ref{sec:dr} and ~\ref{sec:newgal} we will present the techniques, the reduction procedures and the detailed analysis of the two new galaxies observed by MUSE CGCG097-079 and CGCG097-073. Readers interested only in the discussion of the statistical sample of galaxies can move directly to Section~\ref{sec:emlines}.

\section{Observations and data reduction}
\label{sec:dr}
CGCG097-079 and CGCG097-073, together with their ionised gas tails, were observed with MUSE at VLT (UT4) from February 2016 to April 2017 as part of programmes 096.B-0019(A) and 098.B-0020(A) (PI G. Gavazzi), using the Wide Field Mode and the nominal wavelength range of the instrument.
These galaxies were observed at airmass $\approx 1.4$ under clear skies or, occasionally, with thin clouds. The final point spread function (PSF) full-width-at-half-maximum (FWHM) covers the range 0.75-1.00 arcsec with  a median value over the entire mosaic of 0.85 arcsec. Flux calibration has been achieved by means of observations of the standard star GD 153 for 160 seconds.
The mosaic is composed of 8 MUSE pointings covering the optical extent of the two galaxies and their respective tails. 

\begin{figure*}
	\includegraphics[width=0.95\textwidth]{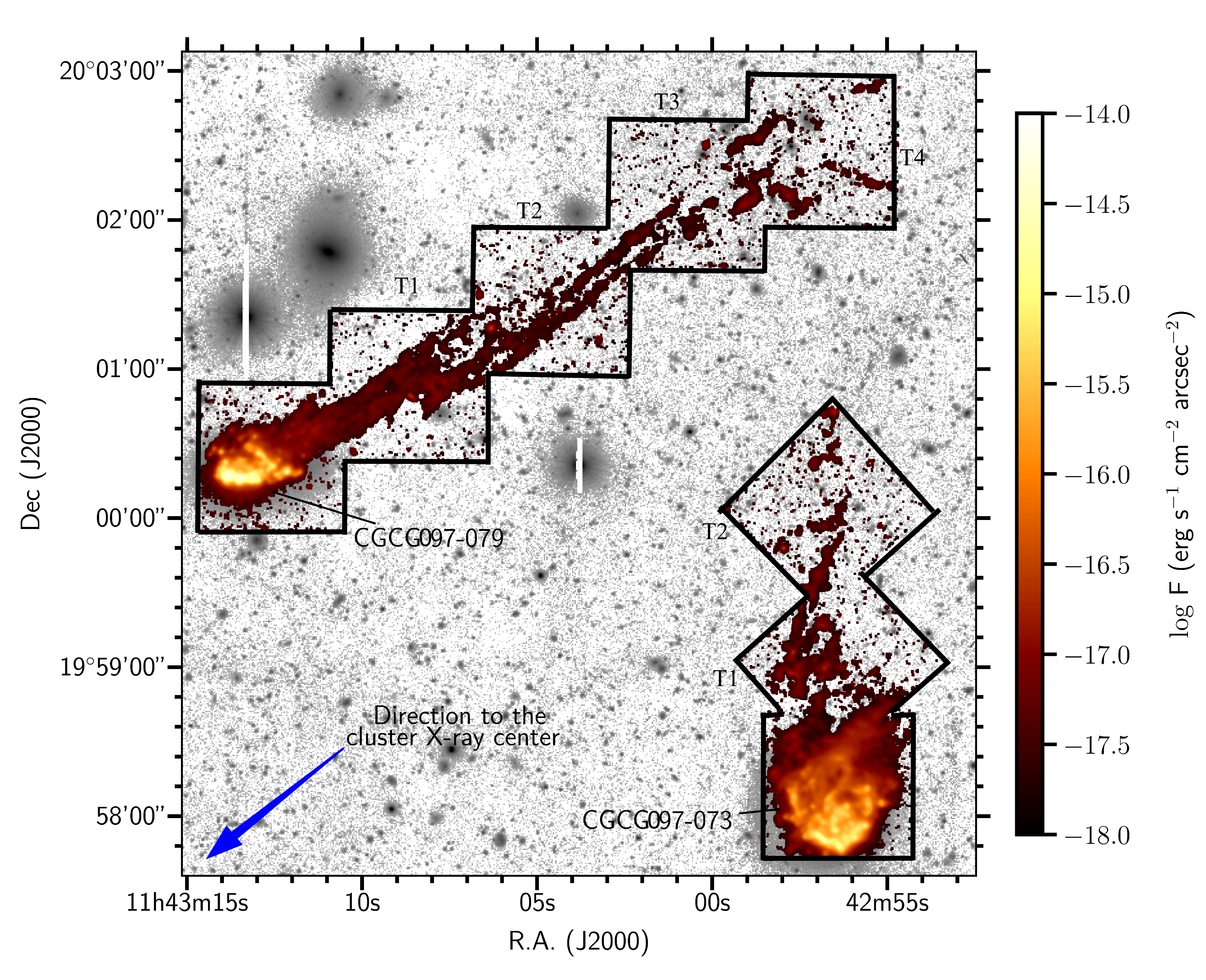}
    \caption{The Subaru $B$-band image of the field covering CGCG097-079 and CGCG097-073. The H$\alpha$ map of the ionised gas from the MUSE datacube is overplotted in red. The black boxes mark the edges of the individual MUSE exposures. The blue arrow indicates the direction to the X-ray center of Abell 1367.}
    \label{fig:SUBARU_MUSE}
\end{figure*}

In particular, two pointings were centered on the galaxy stellar disks, while the other ones on the tails, as shown in Fig.~\ref{fig:SUBARU_MUSE}.
Each observation consisted of two 1290 seconds exposures (or three 777 seconds in the pointings of the tails). Each science pointing is interspersed by a sky exposure of 4 minutes and 2 minutes for the observations taken in 2016 and 2017, respectively. The instrument was rotated by 90 degree after each exposure to improve the final uniformity of the reconstructed image.

The raw data have been processed with the MUSE data reduction pipeline (v2.8.3, \citealt{Weilbacher20}) and more details about the data reduction steps can be found in the previous papers of this series \citep{Fumagalli14,Fossati16,Consolandi17,Fossati19}. These steps include bias subtraction, flat fielding, wavelength calibration and illumination corrections, and the reconstruction of exposures into datacubes. Despite the use of the twilight exposures as calibrators, the spatial uniformity of the reconstructed cubes remain uneven. For this reason we use the internal illumination uniformity calibration (dubbed autocalibration in the data reduction software) which uses background spaxels to individually calibrate the flux level of each stack. This procedure requires a field that is mostly empty of continuum sources and is therefore not applied for the pointings centred on the galaxy disks. 
In order to produce a mosaic containing the 8 different pointings we construct a reference WCS grid that encompasses our observations as described in \citet{Fossati19}, with a 0.2~arcsec spatial pixel scale and a 1.25~\AA\ spectral step from 4750~\AA\ to 9300~\AA. All science exposures are then astrometrically calibrated using SDSS point sources, and projected onto the reference grid using the MUSE pipeline.

We use the ZAP code \citep{Soto16} in order to improve the removal of sky line residuals. This tool is based on principal component analysis (PCA) to isolate and remove from the datacube residual sky features.
It is possible to operate with this code using the sky in the science image or an external one. For our purposes we have to assume that faint emission lines from the target galaxies fill most of the field of view of each exposure. We therefore use the external sky exposures to generate the PCA decomposition that is then applied to the science exposures as also done in \citet{Fossati19}.
Furthermore, after checking the reliability of the photometric calibration of the data also as a function of wavelength, we perform an in-house sky subtraction of the sky continuum which is known to vary between exposures and is not corrected by ZAP. 

Specifically, we manually select small background regions without detectable emission from continuum sources in each cube. Enough background regions are present even in the pointings covering the galaxy disks, thanks to the large FOV of MUSE. We also explicitly tested that the choice of different regions does not affect the results presented here. We coadd the pixels of these regions in each cube to obtain spectra of the sky background for which we derive average values in bins of 150$\AA$ along the wavelength axis. By fitting these values with a fourth degree polynomial function and subtracting each best fit function from the respective data cube, we refine the zero level of the sky subtraction. We stress that this procedure has no effect on the absolute level of the flux of emission lines.

Lastly we combine the exposures with mean statistics and we test the flux calibration of our final mosaic by comparing the $r-$band flux of point-like sources in the field with the SDSS database, finding consistency within $\approx5\%$.

\subsection{Stellar continuum and emission line fitting}
\label{sec:cont}

We start by extracting the flux maps of ionised gas emission lines covered by MUSE in order to characterise the physical properties of the gas in the disks and in the tails of the two galaxies. Before fitting these lines, we apply to the mosaic cube a 10 $\times$ 10 pixels (2 $\times$ 2 arcsec) median filter to improve the signal-to-noise ($S/N$) ratio for each pixel without significantly degrading the original spatial resolution.

Moreover, we perform a stellar continuum correction aimed at removing the underlying Balmer absorption using the GANDALF fitting code \citep{Sarzi06} and a selection of stellar spectra from the the Indo-U.S. Library of Coud\'e Feed Stellar Spectra library \citep{Valdes04}, similarly to what done in \citet{Fossati19}. For this purpose, we run the code only on cutouts centered on the two galaxies  CGCG097-079 and CGCG097-073. Then, for each spaxel, it is possible to subtract the best fit stellar continuum spectrum obtaining a cube of pure emission lines.

As a result of this procedure, GANDALF returns maps of the stellar kinematics for each galaxy \citep{Fossati19}. However, in this paper we present two low-mass galaxies for which it is difficult to obtain reliable stellar kinematics parameters due to the low surface brightness of the stellar emission. We therefore do not use these products in the following analysis.
In order to obtain a new datacube with only the emission lines associated to the two galaxies investigated, we have also to remove isolated background objects from the field.
Four background galaxies at $z \sim$ 0.82 are found from their [OII] and [OIII] emission lines, and are then masked to avoid contamination.

Using the KUBEVIZ code as detailed in previous papers of this series,
we then extract flux maps of the strongest emission lines falling in the MUSE wavelength range, namely:
$\rm H\beta$; [OIII]$\lambda \lambda 4959,5007$; [OI]$\lambda 6300$; [NII]$\lambda\lambda 6548,6583$; H$\alpha$; [SII]$\lambda\lambda 6716,6731$. The procedure fits the entire MUSE datacube flagging spaxels with $S/N > 5$ in H$\alpha$ and with uncertainties on the kinematics smaller than 50 km s$^{-1}$ as robust fits that we use for the analysis.
Due to the very similar redshift of the two galaxies, we assume for the entire mosaic the systemic redshift of CGCG097-079, $z_{sys} = 0.02363$ \citep{Gavazzi10}, as velocity zero point.

\section{Properties of CGCG097-079 and CGCG097-073}
\label{sec:newgal}
In this section, we present an in-depth analysis of the new observations of CGCG097-079 and CGCG097-073. Specifically, we present the results obtained from the study of the gas kinematics and the ionization conditions of the diffuse gas in these two galaxies.

\subsection{Gas kinematics}
\label{sec:gaskin}
\begin{figure*}
	\includegraphics[width=0.95\textwidth]{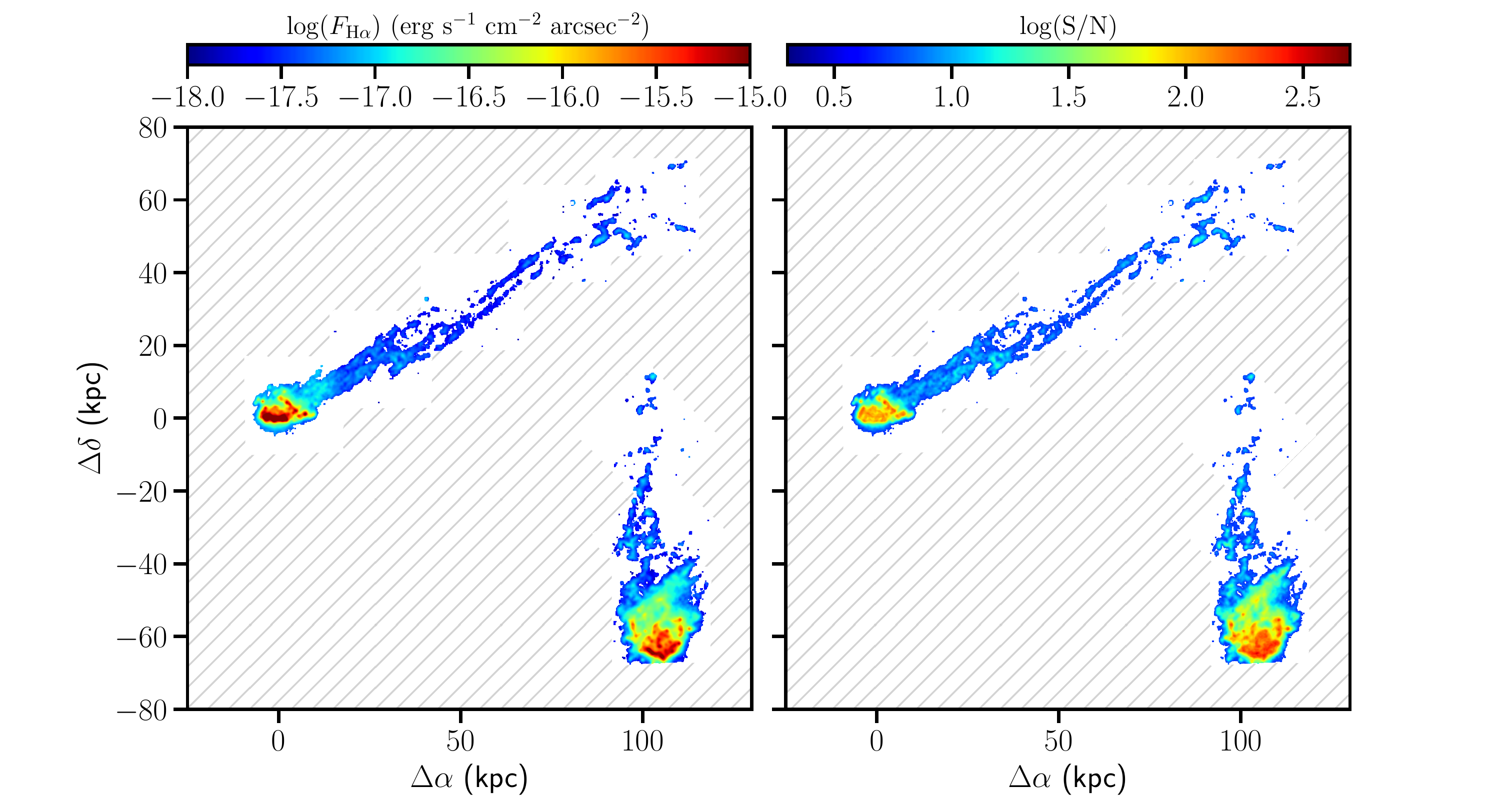}
    \caption{\textit{Left:} H$\alpha$ surface brightness map of the ionised gas in CGCG097-079 and CGCG097-073.  \textit{Right:} $\log S/N$ map obtained dividing the H$\alpha$ flux value by the associated errors in each pixel of the datacube. Areas not covered by MUSE observations are shaded in gray. }
    \label{fig:fluxmap}
\end{figure*}
\begin{figure}
	\includegraphics[width=\columnwidth]{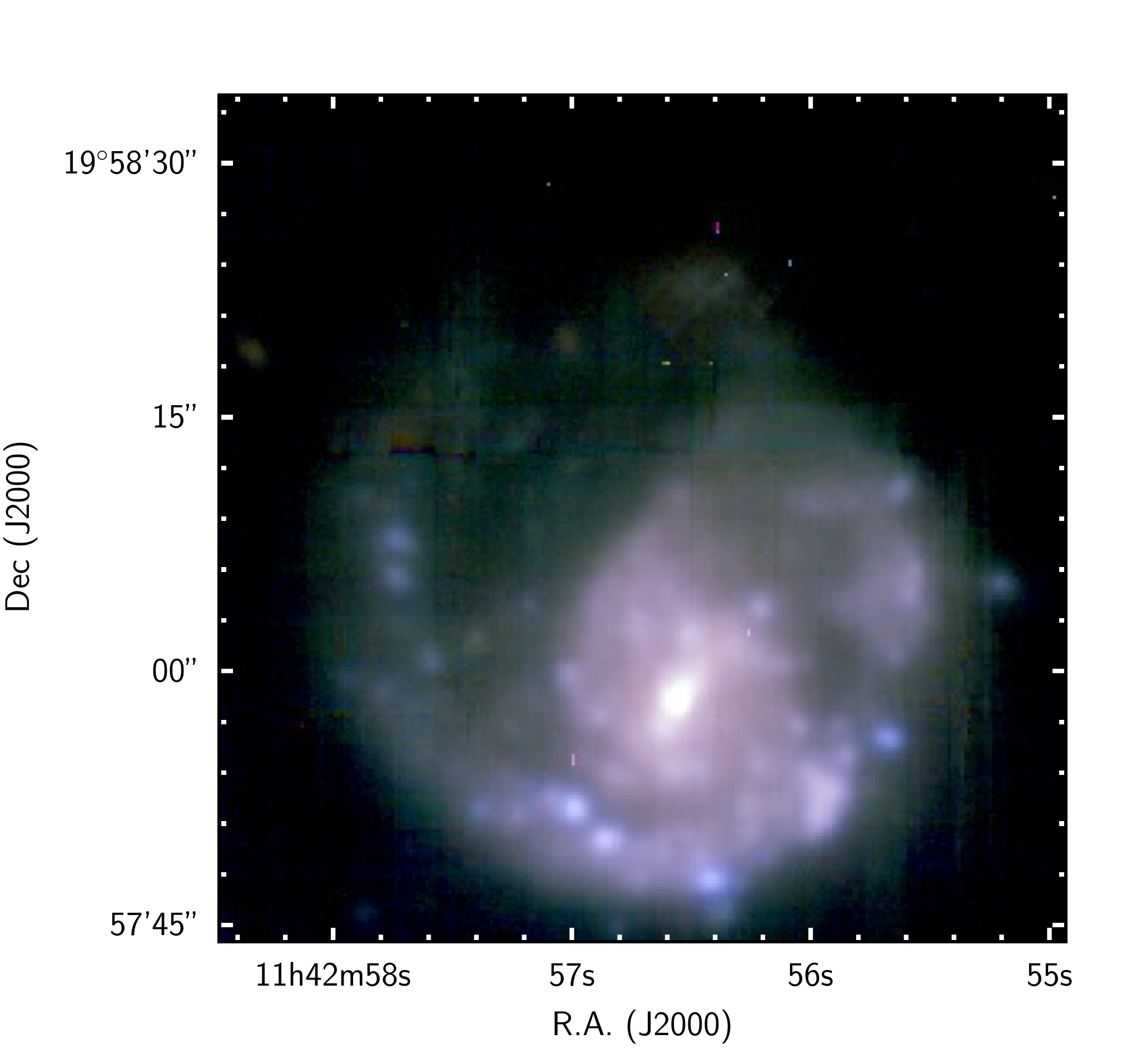}
    \caption{RGB false color of CGCG097-073, obtained by convolving the MUSE data cube with the \textit{gri} SDSS filters.
    }
    \label{fig:73}
\end{figure}
\begin{figure}
	\includegraphics[width=\columnwidth]{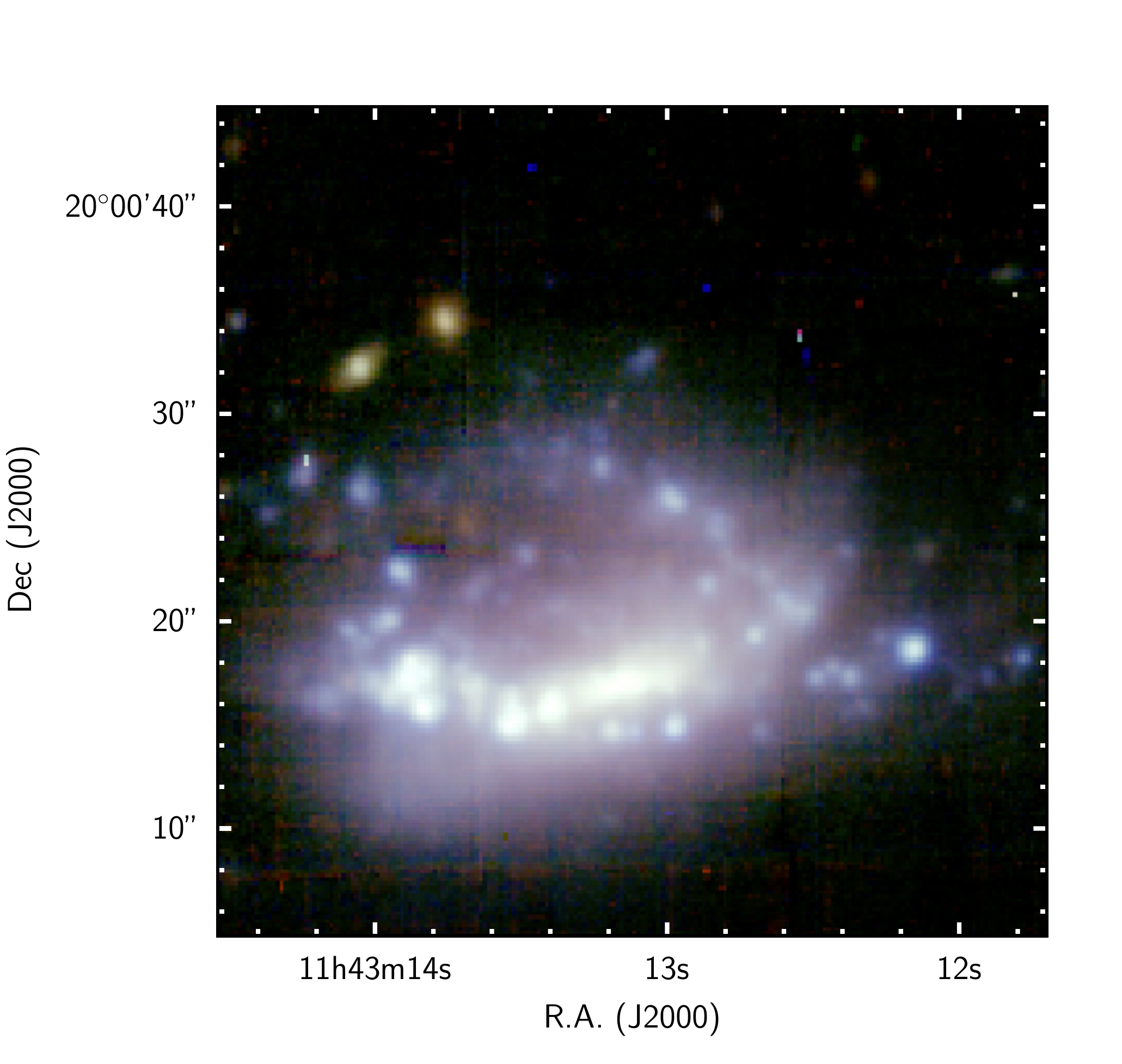}
    \caption{The same of Fig.~\ref{fig:73}, but for CGCG097-079
    }
    \label{fig:79}
\end{figure}

The sensitivity and high spatial resolution of the MUSE observations allow us to study maps of the ionised gas emission and kinematics in detail. In Fig.~\ref{fig:fluxmap} we present the H$\alpha$ map of the observed mosaic. We can immediately see two long filamentary tails starting from the center of the two galaxies. In particular, for CGCG097-079 the tail extends in projection for about 100 kpc with surface brightness approximately ranging from 10$^{-18}$ erg s$^{-1}$ cm$^{-2}$ arcsec$^{-2}$ to 10$^{-17}$ erg s$^{-1}$ cm$^{-2}$ arcsec$^{-2}$, which are typical of ionised gas RPS tails  \citep{Yoshida12, Boselli16B, Yagi17, Consolandi17}.
Similarly, CGCG079-073 presents a tail extension of about 70 kpc and roughly the same surface brightness of the other galaxy tail.
The morphology of the galaxies and the surface brightness are consistent with what measured in narrow-band imaging by \citet{Yagi17}.
The strongest emission in H$\alpha$ is located at the leading edge of the two galaxies and despite an intense star formation activity in the two galaxy disks (that we can clearly observe in Fig.~\ref{fig:73} for CGCG097-073 and Fig.~\ref{fig:79} for CGCG097-079) we only observe a few compact regions (putative HII regions) in the low surface brightness tails. 
Following \citet{Kennicutt98B}, we estimate the SFR from the integrated H$\alpha$ flux within the galaxy disks. We find $\approx 1.62 \pm 0.05~\rm M_{\odot}~yr^{-1}$ and $\approx 1.20 \pm 0.02 \rm M_{\odot}~ yr^{-1}$, respectively for CGCG097-079 and CGCG097-073, after correcting for dust attenuation using the Balmer decrement in each pixel of the galaxy disks. Here we have made the assumption that the gas ionization is dominated by photoionization (and not by shocks or AGN), and we verified that this assumption holds as we will show in Figure \ref{fig:bpt}.

\begin{figure*}
	\includegraphics[width=0.95\textwidth]{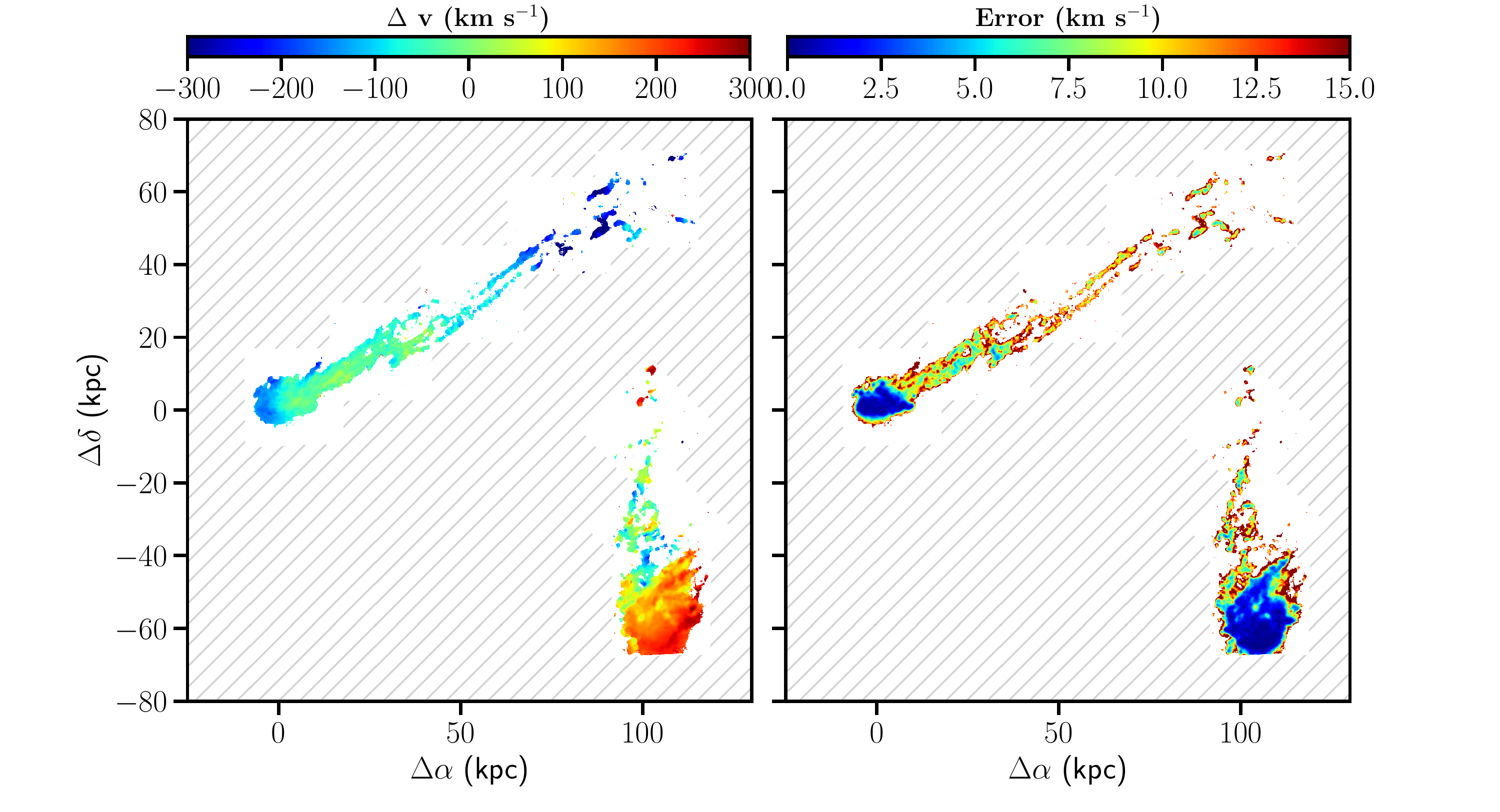}
    \caption{\textit{Left:} Velocity map of the H$\alpha$+[NII] line complex relative to the galaxies systemic redshift. \textit{Right:} Error map of the velocity. Areas not covered by MUSE observations are shaded in gray.}
    \label{fig:velmap}
\end{figure*}

\begin{figure*}
	\includegraphics[width=0.95\textwidth]{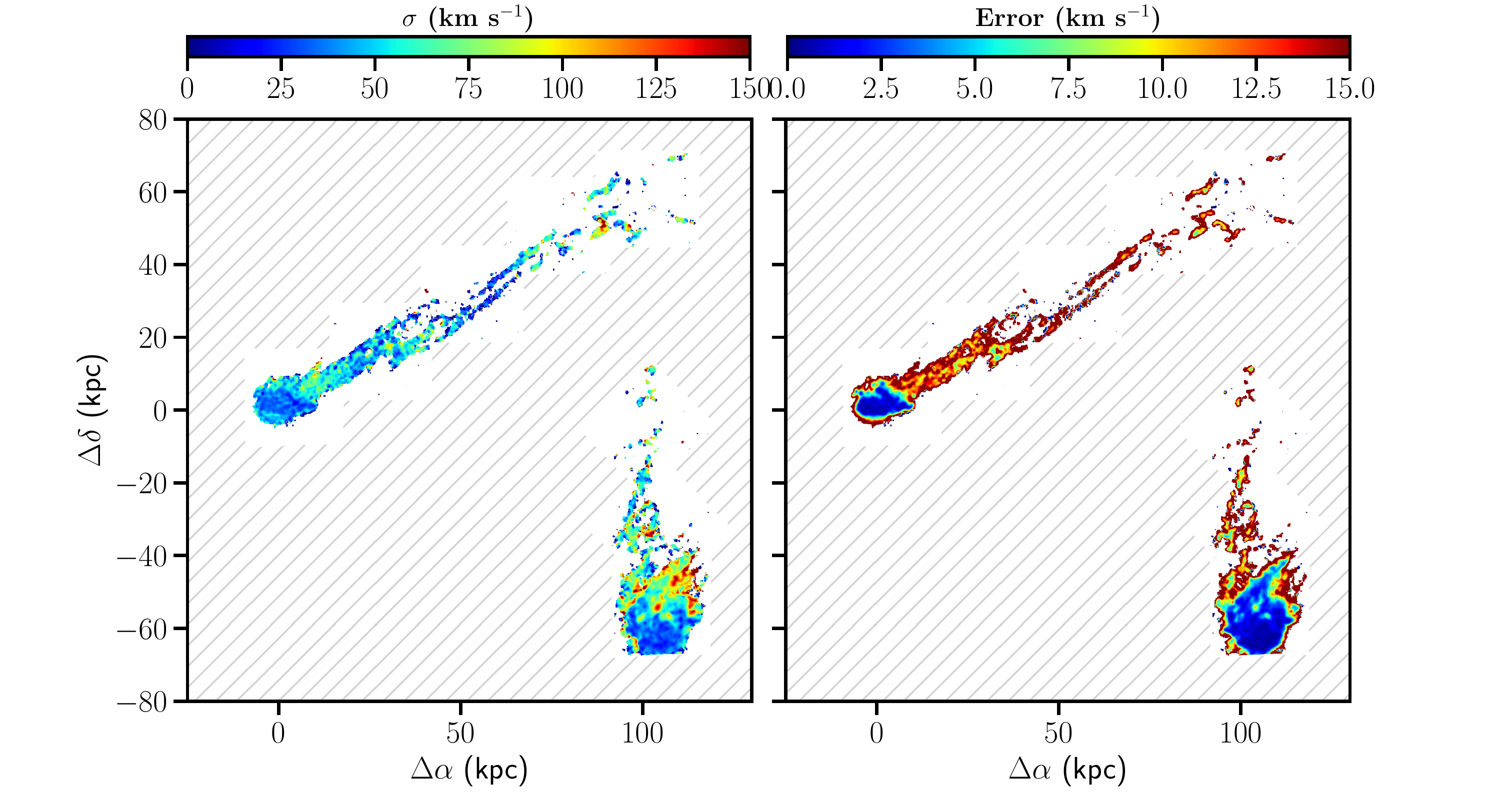}
    \caption{\textit{Left:} Velocity dispersion map of the H$\alpha$+[NII] line complex. \textit{Right:} Error map of the velocity dispersion. Areas not covered by MUSE observations are shaded in gray.}
    \label{fig:sigmap}
\end{figure*}

Maps of the velocity relative to systemic redshift, $\Delta v = v - v_{sys}$, and the intrinsic velocity dispersion $\sigma$ from the fit of the H$\alpha$+[NII] line complex are shown in Fig.~\ref{fig:velmap} and Fig.~\ref{fig:sigmap}.
From the velocity map, we see ordered rotation for the two galaxies and we observe that CGCG097-073 has a larger recessional velocity along the line of sight compared to the other galaxy. Furthermore, we see a clear velocity gradient in the tail of CGCG097-079, in the direction to the cluster X-ray center (Fig.~\ref{fig:SUBARU_MUSE}), suggesting that the gas at larger distance from the galaxy is decelerated to lower velocities by the ICM ram-pressure (see e.g. \citealt{Tonnesen10} for numerical simulations of this effect). 

Moreover, one can notice that in CGCG097-073 there is a side tail pointing in the NW direction (identified as T1 W in Figure \ref{fig:reg}) presenting slightly higher values of velocity dispersion. However, further discussions about emission line ratio maps in the following section suggests that turbulence in this region does not appear to be associated with a variation in the gas ionization conditions, therefore we propose it to be a purely dynamical process observed in projection on the sky.
Moreover, it is important to point out that velocity dispersion values below 50 km s$^{-1}$ are marginally resolved at the MUSE resolution \citep[e.g.][]{Boselli21} and consequently are more uncertain and affected by systematics in the estimate of the line spread function.

\subsection{Emission line diagnostic}
\label{sec:emlines7973}
In the previous section we have examined the properties of the H$\alpha$+[NII] emission. We now turn to the study of the physical properties of the ionised gas using line ratios. \citet{Kewley01, Kewley06, Ferland08, Rich11} have shown that different ionization processes (photo-ionization from hot stars or from harder sources, shocks, turbulence, and magneto-hydrodynamic waves) lead to different ratios in the intensity of the emission lines.  In order to place constraints on which ionization processes are taking place in these galaxies, we make use of three optical diagnostic diagrams (BPT diagrams; \citealt{Baldwin81}). Specifically, we show the ratio between [OIII]$\lambda$5007/H$\beta$ and either [NII]$\lambda$6583/H$\alpha$ or [OI]$\lambda$6300/H$\alpha$ or [SII]$\lambda\lambda$6716, 6731/H$\alpha$.

\begin{figure*}
	\includegraphics[width=0.95\textwidth]{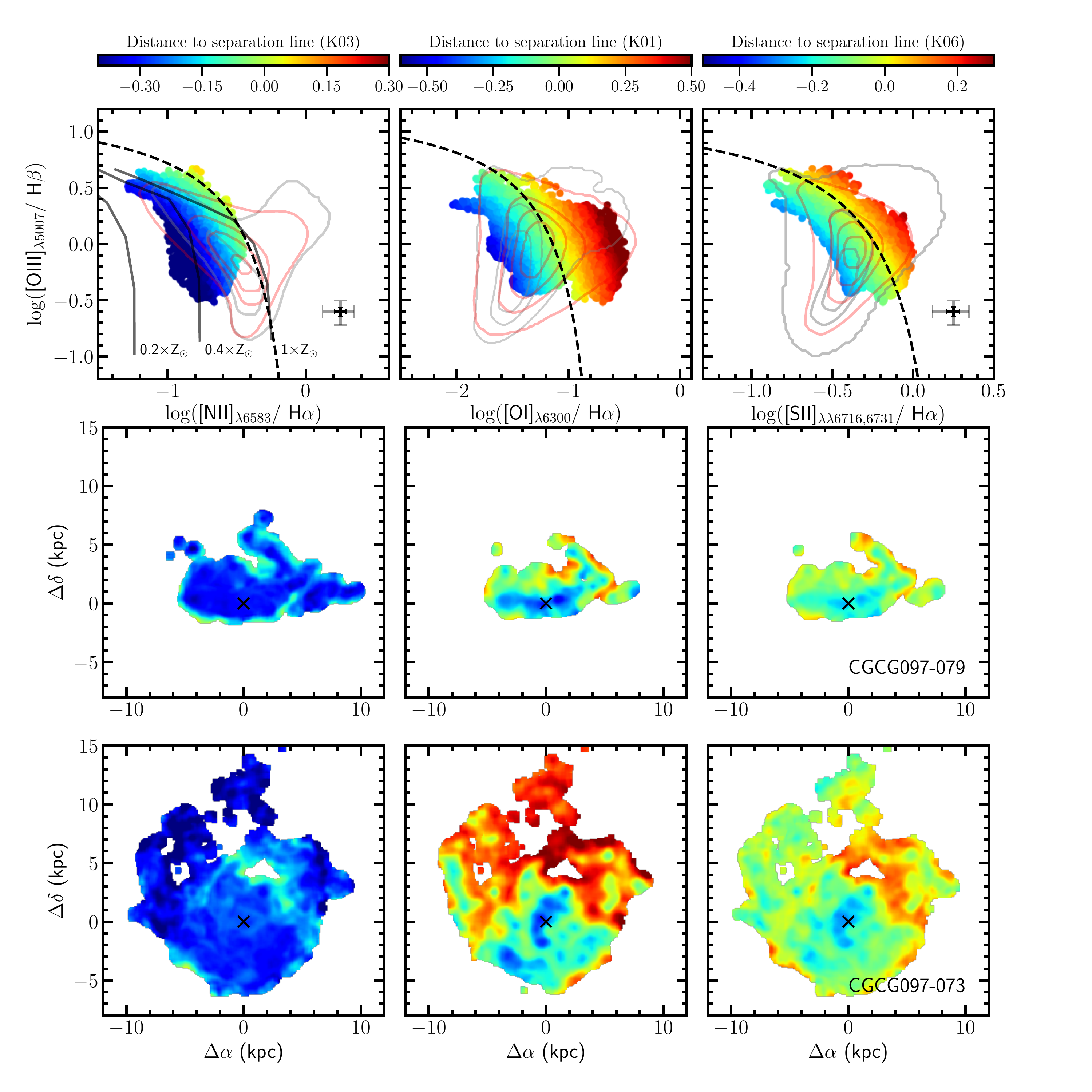}
    \caption{\textit{Top row}: BPT diagnostic diagrams [OIII]/H$\beta$ vs. [NII]/H$\alpha$ (left), [OI]/H$\alpha$ (centre), [SII]/H$\alpha$ (right) derived from cutouts of the MUSE mosaic centred on CGCG097-073 and CGCG097-079. We show only spaxels with $S/N > 5$ for all emission lines considered. The color bars represent the distance from the dashed lines which separate stellar photo-ionization processes (left of the lines) from AGN/shock ionization (right of the lines) as defined in \citet{Kauffmann03} (K03, left), in \citet{Kewley01} (K01, center) and in \citet{Kewley06} (K06, right). The black and gray crosses show the characteristic error of line ratios with $S/N = 15$ and $S/N = 5$, respectively. The black solid lines in the top-left panel show the tracks from photo-ionization models of \citet{Kewley01} for three different metallicities (0.2, 0.4, 1 Z$_{\odot}$). The grey contours are obtained from the nuclear spectra of a sample of 50,000 SDSS galaxies at $0.03 < z < 0.08$ whose emission lines are detected at $S/N>5$.The pink contours are obtained from spaxels in the outer disks of galaxies ($R>1~R_e$) from the MaNGA DR17 sample that satisfy the $S/N$ cut above. The innermost to outermost contours include the 25\%, 50\%, 75\% and 98\% of each dataset, respectively.
    \textit{Middle and bottom row}: The spatial (on-sky) position of the spaxels colour coded as in the top row panels is shown for CGCG097-079 and CGCG097-073 in the middle and bottom row panels, respectively. The black crosses indicate the optical center of the galaxies.}
    \label{fig:bpt}
\end{figure*}

When considering only spaxels with $S/N>5$ for each emission line contributing to a given BPT diagram we are left only with spaxels on the galaxy disks of CGCG097-079 and CGCG097-073 and their immediate surroundings where the emission line flux is high enough to obtain robust line ratios in individual spaxels. 
We present these results in Fig.~\ref{fig:bpt}. The top panel shows the distribution of individual spaxels in the aforementioned BPT diagrams. Each spaxel is color coded by its distance from the lines separating the region dominated by photo-ionization from the one dominated by active galactic nuclei (AGN) or shocks, following the separation of \citet{Kauffmann03}. 
The distributions of line ratios from nuclear spectra of a sample of 50,000 SDSS galaxies at $0.03 < z < 0.08$ whose emission lines are detected at $S/N>5$ \citep{Fossati16} are shown with grey contours. The innermost to outermost contours include the 25\%, 50\%, 75\% and 98\% of the full dataset, respectively. However, this SDSS sample only represents the nuclear properties of galaxies which might be different from their outskirts which are likely providing most of the stripped gas. To overcome this possible issue we add to the diagnostic diagrams the same density contours from a sample of individual spaxels from the DR17 of the Mapping Nearby Galaxies at APO (MaNGA) IFU survey of $\sim 10,000$ galaxies \citep{Abdurrouf21}. For each BPT diagram, we selected spaxels outside the effective radius ($R>1~R_e$) of each individual galaxy whose emission lines are detected at $S/N>5$. This sample includes roughly half a million spaxels in the BPT diagram using the fainter [OI] line and roughly 2 million spaxels in the other two BPT diagrams. Density contours from this sample (at the same levels mentioned above) are shown in pink in Fig.~\ref{fig:bpt}.
We also verified that by using spaxels from the entire galaxy disks the density contours would appear very similar to the SDSS sample, due larger weight given to the inner parts of the disks. Despite the differences seen in these two reference samples, the BPT diagnostic made using [OI]/H$\alpha$ shows enhanced [OI]/H$\alpha$ in the RPS tails that appear not to be typical neither of the galaxy nuclei nor of their disk outskirts. We will discuss this point in a more quantitative fashion in Sec.~\ref{sec:filament}. In the middle and bottom panels we show the spatial distribution of the spaxels from the two galaxies, with points color-coded as above.

Fig.~\ref{fig:bpt} shows that only the BPT diagram with [OI]$\lambda$6300/H$\alpha$ shows significant deviations from the photo-ionization region, with a large fraction of points downstream of the galaxy CGCG097-073 exhibiting a significantly enhanced [OI]/H$\alpha$ ratio. This line ratio also appears to be significantly enhanced in the inner part of the RPS tail for CGCG097-079 as can be seen in the bottom panel of Figure \ref{fig:bpt}. 
Indeed this ratio has been typically considered as the most sensitive tracer of shocks and turbulence in RPS tails \citep{Yoshida12, Fossati16, Consolandi17, Poggianti18, Boselli18b, Fossati19, Poggianti19}. Conversely, [SII]/H$\alpha$ is only marginally enhanced in the surroundings of the galaxy disks, and only slightly deviates from the photo-ionization sequence. Lastly, the ratio [NII]/H$\alpha$ is entirely consistent with stellar photo-ionization although the ratio is also known to depend on the metallicity of the galaxy, with lower metallicity objects having an intrinsically lower [NII]/H$\alpha$. In particular, the lower envelope of [NII]/H$\alpha$ values for our two galaxies (as seen from the top left panels) suggests a sub-solar metallicity for these objects, as one could expect from the low stellar mass of these galaxies \citep{Tremonti04}. This feature is also confirmed by the metallicity vs stellar mass relation from \citet{Maiolino19} which gives typical values of $\sim 0.3-0.8~\rm  Z_{\odot}$ for galaxies with stellar mass of $10^9-10^{10}$ M$_{\odot}$.

\begin{figure}
	\includegraphics[width=\columnwidth]{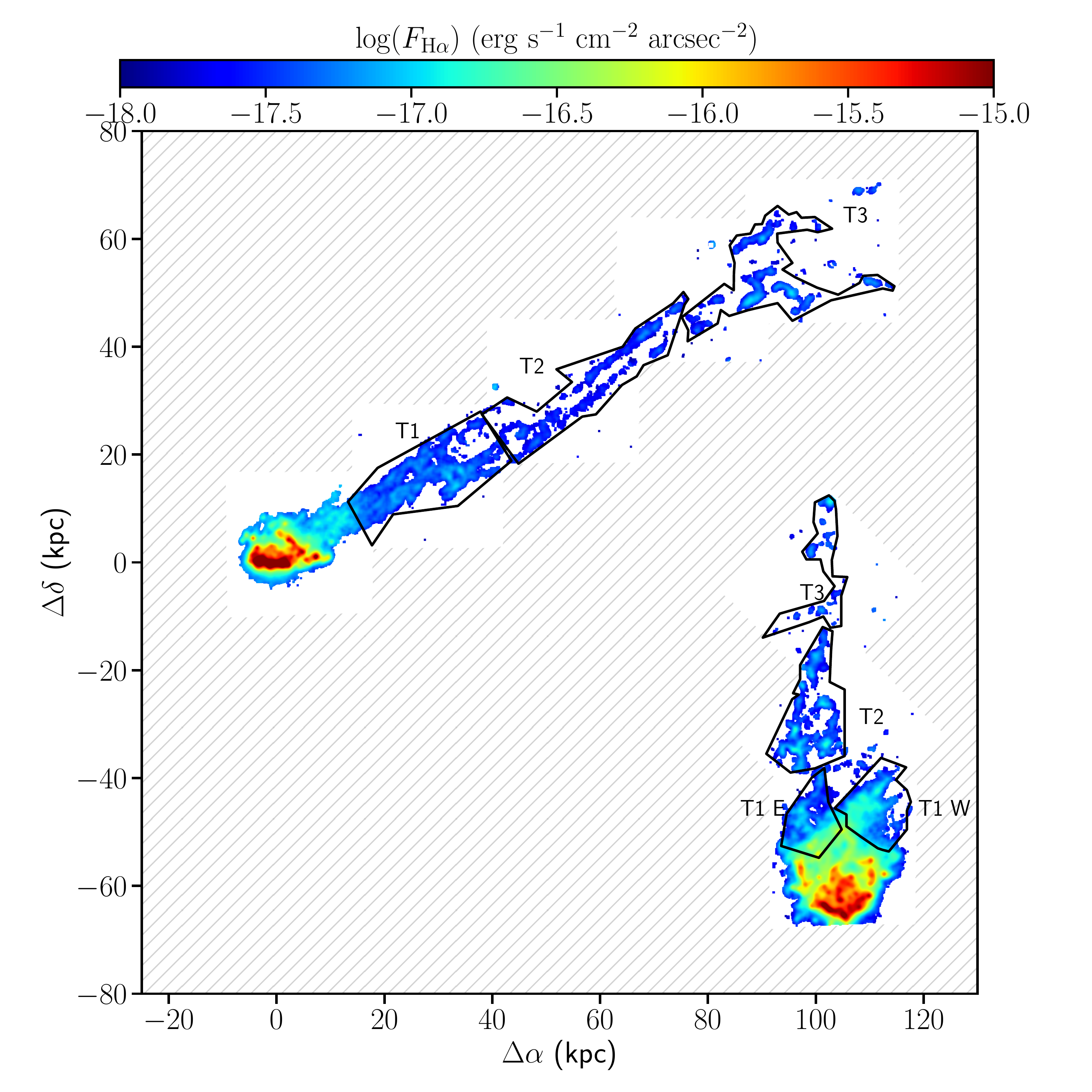}
    \caption{H$\alpha$ map of the ionised gas in CGCG097-079 and CGCG097-073. Black polygonal apertures represent selected regions used to derive the composite spectra as outlined in the text. Specifically, in CGCG097-079, \textit{T1}, \textit{T2} and \textit{T3} extend from the nearest area to the galaxy disk to the furthest. The same is true for \textit{T1}, \textit{T2} and \textit{T3} in CGCG097-073 except for \textit{T1 E} and \textit{T1 W} which are roughly at the same distance from the galaxy but have a different ionised gas surface brightness and kinematics. Areas not covered by MUSE observations are shaded in gray.}
    \label{fig:reg}
\end{figure}

A resolved, pixel-by-pixel analysis of the BPT diagrams is possible only in the central regions of the galaxies due to the high $S/N$ of the emission lines involved. To study the processes taking place in the filamentary tails of the galaxies in detail, we generate instead high $S/N$ composite spectra. For this purpose, we focus on three and four independent regions (see Fig.~\ref{fig:reg}) for CGCG097-079 and CGCG097-073, respectively. These regions cover the RPS tails of the two galaxies at increasing distances from the galaxy disks. The closest region to the disk of CGCG097-073 is split in two parts to evaluate the potential difference in the line ratios of these two regions which appear as distinct both in the ionised gas surface brightness and its kinematics.
To extract the co-added spectra, we follow \citet{Fossati16} and \citet{Fossati19}, and we select the spaxels in each region where $S/N(\rm H\alpha)>5$. We then sum the spectra of these spaxels after shifting them to rest-frame wavelengths using the kinematics information. Then, for each region, we fit the composite spectra with KUBEVIZ in order to obtain emission line fluxes. As discussed in detail in Section \ref{sec:filament}, the spectra extracted from the tails prominently show enhanced [OI]/H$\alpha$, and typical values of other line ratio diagnostics as seen in previous studies of RPS galaxies.

Having discussed the properties of these two new galaxies, in the following section we turn to the properties of the larger sample, and put these results in the broader context of galaxies suffering RPS in A1367.

\section{The ram-pressure stripping scenario in A1367}
\label{sec:emlines}

With the goal of obtaining a more complete view of ram-pressure stripping in A1367 and in other massive clusters, we now combine the new observations of CGCG097-079 and CGCG097-073 presented in the previous sections of this paper with other objects studied in this series of papers to build a sample of known RPS galaxies. We can proceed then to study the properties of the diffuse gas filaments around these galaxies (Section~\ref{sec:filament}) and the regions of interaction between the disks and the ICM (Section~\ref{sec:interface}).

\subsection{Ionization conditions outside galaxies}
\label{sec:filament}

This analysis on the diffuse gas in the tails of several galaxies with different properties (e.g. stellar mass, metallicity, star formation activity) is the first step towards a deeper knowledge of the interaction between the cluster ICM and the galaxy ISM. 

\begin{figure*}
	\includegraphics[width=0.95\textwidth]{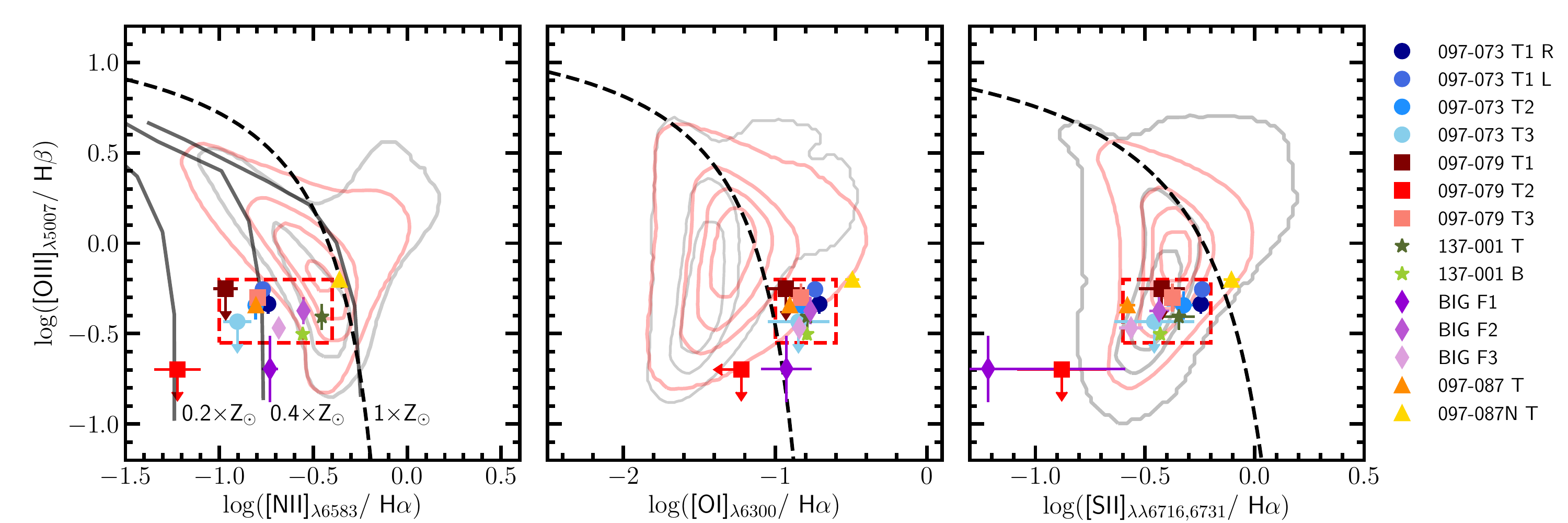}
    \caption{BPT diagnostic diagrams [OIII]/H$\beta$ vs. [NII]/H$\alpha$ (left), [OI]/H$\alpha$ (centre), [SII]/H$\alpha$ (right) derived from the composite spectra extracted from the seven regions of CGCG097-073 and CGCG097-079 presented in Fig.~\ref{fig:reg} and gas diffuse filaments from BIG, CGCG097-087, CGCG097-087N and ESO137-001 obtained from the previous papers of this series. Error bars are obtained from KUBEVIZ fits. Downward or leftward arrows represent the 2$\sigma$ upper limits of the line ratios when metal lines are too faint to be detected. The grey and pink contours, and the solid and dashed black lines are as in Fig.~\ref{fig:bpt}.}
    \label{fig:1Dbpt}
\end{figure*}

Fig.~\ref{fig:1Dbpt} shows the BPT diagrams obtained from the co-added spectra of different regions outside galaxies  and on the RPS tails in our sample. Specifically, we show seven regions of CGCG097-079 and CGCG097-073 (Fig.~\ref{fig:reg}), tail (T) and blob (B) regions in ESO 137-001 in the Norma Cluster \citep{Fumagalli14, Fossati16},  regions F1, F2 and F3 in the Blue Infalling Group (BIG) \citep{Sakai02, Cortese06, Fossati19}, and finally the ionised gas tails in CGCG097-087 and CGCG097-087N \citep{Consolandi17}.
While the Balmer H$\alpha$ and H$\beta$ lines are robustly detected in each composite spectrum, metal lines can be too faint to be detected in some regions. In these cases, we show the 2$\sigma$ upper limit of the line ratios as downward or leftward arrows.

The [OIII]/H$\beta$ vs [OI]/H$\alpha$ diagram confirms the result found also in Fig.~\ref{fig:bpt}. Indeed, almost all points populate the bottom-right corner of the diagram, which we interpret as an indication of the relevance of ionization from shocks induced by the RPS events. The reference contours from SDSS nuclear regions (grey contours) or MaNGA outer disks (pink contours) indicate that these line ratios are not typical in local star-forming galaxies.

Furthermore, we note that the [OIII]/H$\beta$ vs [SII]/H$\alpha$ diagram is not a particularly good discriminator for the conditions in the RPS tails, as all the regions examined here are consistent with both reference samples of unperturbed star-forming galaxies. Lastly, the [OIII]/H$\beta$ vs [NII]/H$\alpha$ diagram indicates typical [NII]/H$\alpha$ values consistent with photo-ionization from HII regions, as mentioned in the previous section. In this plane, the points present a larger scatter compared to the other diagrams, a feature which we ascribe to the metallicity sensitivity of this indicator. If used alone, this latter diagram would let us conclude that standard photoionization conditions prevail in RPS tails, while the peculiar [OI]/H$\alpha$ ratios indicate a significant contribution of other ionization processes.

An outlier with respect to the other points is the CGCG097-087N, marked by a yellow triangle in the BPT diagrams. We notice in fact that this region shows an higher value compared to the other galaxies in all ratios on the x-axis of the BPT diagrams: [NII]/H$\alpha$, [OI]/H$\alpha$, and [SII]/H$\alpha$. This is probably related to the tidal interaction driven by its higher mass companion CGCG097-087, as already discussed in \citet{Consolandi17}. The regions labelled BIG F1 and CGCG097-079 T2 are also outlying having lower values (or upper limits) on the [OIII]/H$\beta$ ratio. This could indicate a weak ionization parameter but could also be a consequence of larger errors arising from shallower observations in these regions, or some skyline residuals.

To further quantify the unusual [OI]/H$\alpha$ signature of RPS tails compared to galaxy disks we have identified a rectangular region (red dashed rectangle in Fig.~\ref{fig:bpt}) in each BPT diagram that includes 11 of our 14 datapoints. We exclude the three outliers (CGCG097-087N, BIG F1, and CGCG097-079 T2) for the reasons described above. Each rectangle therefore includes $79\% \pm 11\%$ of our sample. The fraction of MaNGA spaxels, representative of the outer disks of galaxies, that falls in each rectangle is 20\%, and 27\% for the [NII] and [SII] BPT diagrams, respectively. While the difference is significant, it can be ascribed to the limited metallicity and ionization parameter range of our sample. The same fraction of MaNGA spaxels, however, becomes a mere 2\% in the [OI] diagnostic diagram, i.e. there are virtually no galaxy disks in the region of the [OI] BPT diagram where the vast majority of RPS tails live. As a last test we limit the fraction of MaNGA spaxels to those that have  $-0.55<\log({\rm [OIII]/H}\beta)<-0.20$ ratios as low as the RPS tails. In this case we are answering the question, how statistically significant is the [OI]/H$\alpha$ elevated ratio among normal galaxies at fixed [OIII]/H$\beta$? The fraction of MaNGA spaxels raises only to 7\%, roughly an order of magnitude lower than for RPS tails. While selection effects are present even in the MaNGA sample, we are therefore confident that the specific locus occupied by RPS tails in the [OI] BPT diagnostic is not directly related to the properties of the gas ablated from the outer regions of galaxies but that it rather encodes more information on the physics of the stripping process.

\subsection{The RPS interface regions}
\label{sec:interface}

In the previous section we have shown how the full sample shares common physical properties of the diffuse gas in the tails as inferred from the analysis of emission lines through BPT diagrams.
Focusing instead on the RPS interface regions, i.e the regions of interaction between the galaxy disk and the cluster ICM, it is possible to notice that different galaxies show different physical properties from the study of the emission line ratios.
Specifically, we examine the [OIII]/H$\beta$, [NII]/H$\alpha$, [OI]/H$\alpha$, and [SII]/H$\alpha$ ratios to explore in more detail the ionization strength, metal content and contribution of shocks in this interface region. We exclude from the analysis CGCG097-087 and CGCG097-087N because of the complex kinematics due to the existence of multiple velocity components along the line of sight in CGCG097-087 and the gravitational and tidal interactions in CGCG097-087N \citep{Consolandi17}.

\begin{figure*}
    \centering
	\includegraphics[width=1.05\textwidth]{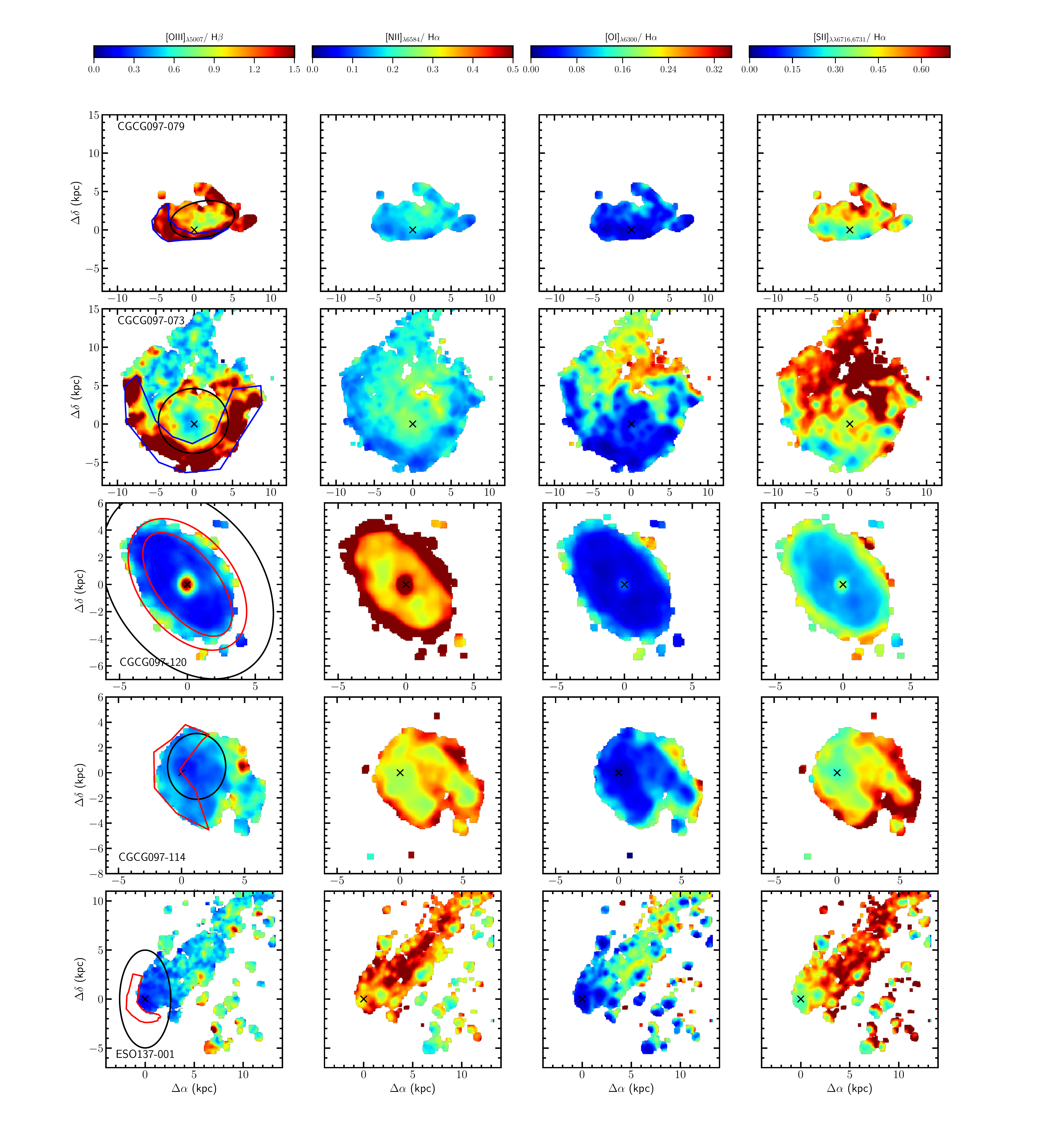}
    \caption{From left to right: [OIII]/H$\beta$, [NII]/H$\alpha$, [OI]/H$\alpha$ and [SII]/H$\alpha$ line ratio maps for CGCG097-079, CGCG097-073 CGCG097-120, CGCG097-114 and ESO137-001 (top to bottom). For visualization purpose, we show only spaxels with $S/N > 3.5$ for all emission lines considered. Black ellipses mark the stellar disks (see text for details). Blue polygonal regions (red for CGCG097-120, CGCG097-114 and ESO137-001 for visualization purpose) are the RPS interface regions extracted from the line ratios. The black crosses indicate the optical center of the galaxies.}
    \label{fig:ratiomap}
\end{figure*}

In Fig.~\ref{fig:ratiomap} we present line ratio maps of five galaxies: CGCG097-079, CGCG097-073, CGCG097-120, CGCG097-114 and ESO137-001, ordered from  top to bottom. The black ellipses in the left panels represent the stellar disk of the five galaxies obtained at a SB limit of 25 mag arcsec$^{-2}$ from $r$-band images of the galaxies extracted from the MUSE data cube. The polygonal regions (blue: CGCG097-079 and CGCG097-073, red: CGCG097-120, CGCG097-114 and ESO137-001) in the same panels mark the projected areas in which galaxies are likely to be interacting with the ICM. They are selected to be on the opposite side of the tails, at the edges of the H$\alpha$ disks, where we can obtain good signal to noise on the emission lines and without a pre-selection on emission line ratios. In ESO137-001 this region is defined as ``front'' in \citet{Fossati16} and we show it for consistency with that work, although the low S/N of lines fainter than H$\alpha$ limits our ability to show a spatially resolved map. The integrated spectrum of the region, however, provides a better S/N on most of the lines we use for the BPT diagnostics.

Focusing first on CGCG097-079 and CGCG097-073 we witness a clear increase of the [OIII]/H$\beta$ ratio in the RPS interface regions. The same areas are also peculiar for the low values of [NII]/H$\alpha$, [OI]/H$\alpha$ and [SII]/H$\alpha$ compared to the rest of the disk. This is consistent with a photo-ionization region with an elevated star formation and a high ionization parameter of the gas \citet{Fossati16}. Indeed, a local increase in specific SFR is usually connected with an increase also in the strength of the [OIII] emission line \citep{Kennicutt92}.
This is the first time in this series of papers in which we identify such a significant increase of [OIII]/H$\beta$ in these regions. Conversely, in CGCG097-120, CGCG097-114 and ESO137-001 we observe an enhanced  [NII]/H$\alpha$ ratio and low values of [OIII]/H$\beta$. In particular, in CGCG097-120 the increase of [NII]/H$\alpha$ ratio is located at the edge of the truncated ionised gas disk (red region of Fig.~\ref{fig:ratiomap}). This is well within the stellar disk and corresponds to the region where the stripping process is currently taking place. A similar increase in [NII]/H$\alpha$ is present at the center of the galaxy where, however, a weak AGN is present. Although only marginally visible in these maps, the integrated spectrum of the interface region of CGCG097-120 also shows elevated [OI]/H$\alpha$ and [SII]/H$\alpha$ compared to the inner regions of the disk.
Similar features are present in the maps of CGCG097-114, although We caution that the interpretation of the higher value of [NII]/H$\alpha$ suffers from a larger uncertainty on the line ratios, which are measured at $3.5<S/N<5$.

Fig.~\ref{fig:ratiomap} gives us a spatially resolved view of the processes taking place at the ICM interface regions. Indeed, we see that different galaxies in our sample can be divided in two macro-branches: galaxies with high [OIII]/H$\beta$ and low [NII]/H$\alpha$ (and with a similar behaviour if [NII] is replaced by [OI] or [SII]) or low [OIII]/H$\beta$ and relatively high [NII]/H$\alpha$. We now investigate whether these different line ratios can be related to different stages of evolution of the RPS process.
To examine in more details the physical conditions of these galaxies and to obtain a more precise picture of this scenario, we generate high $S/N$ composite spectra of the RPS interface regions (blue and red polygonal areas in  Fig.~\ref{fig:ratiomap}) in the same way as in Section ~\ref{sec:filament}.

\begin{figure*}
	\includegraphics[width=0.95\textwidth]{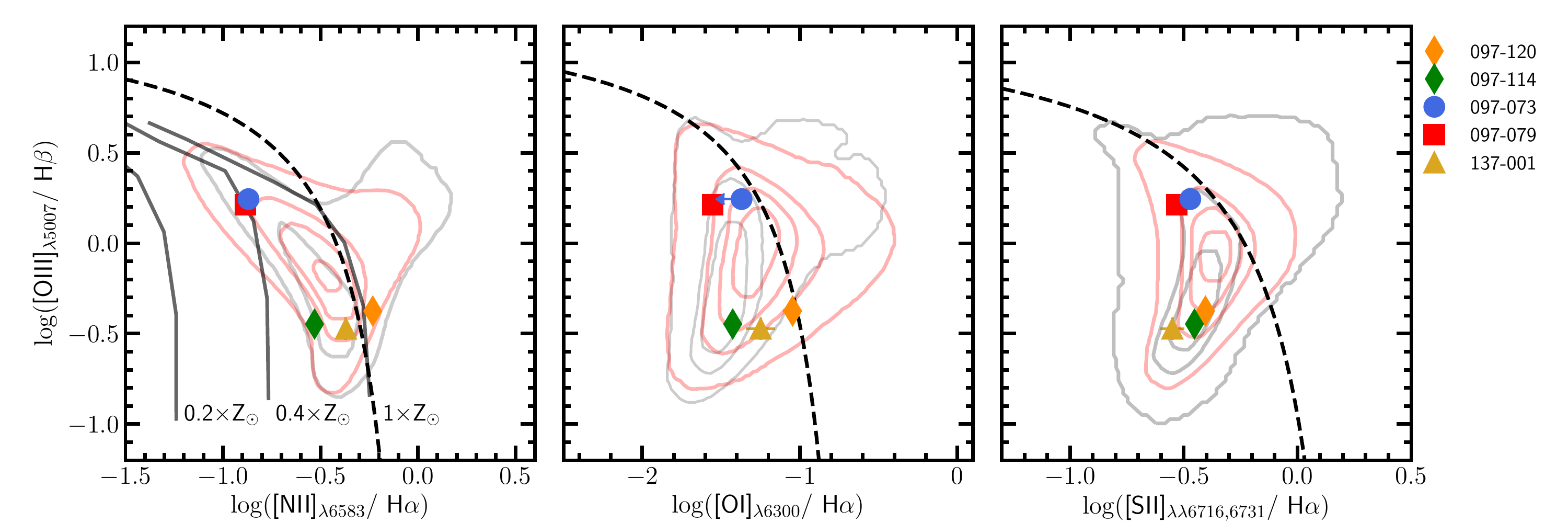}
    \caption{BPT diagnostic diagrams [OIII]/H$\beta$ vs. [NII]/H$\alpha$ (left), [OI]/H$\alpha$ (center), [SII]/H$\alpha$ (right) derived from the composite spectra extracted from the five RPS interface regions of CGCG097-079, CGCG097-073, CGCG097-120, CGCG097-114 and ESO137-001 shown in Fig.~\ref{fig:ratiomap} (blue/red polygonal regions). Error bars are obtained from the KUBEVIZ fit and leftward arrow in the central panel represents the 2$\sigma$ upper limit for [OI]/H$\alpha$. The grey and pink contours, and the solid and dashed black lines are as in Fig.~\ref{fig:bpt}.}
    \label{fig:1Dbptgal}
\end{figure*}

The BPT diagrams for these five regions are shown in Fig.~\ref{fig:1Dbptgal} in which we present the five galaxies described above. Focusing on the left panel, CGCG097-073 (blue circle) and CGCG097-079 (red square) populate the top left region of the diagram corresponding to the top left area of the SDSS/MaNGA samples of galaxies which are populated by objects with strong photoionization from HII regions. 
CGCG097-120 (orange diamond), CGCG097-114 (green diamond) and ESO137-001 (yellow triangle) settle in the central region of the BPT diagram and only CGCG097-120 is at the right of the separation line. A less strong separation along the x-axis can be found in the central and left panels which indicate that these regions are more consistent with each other in terms of [OI]/H$\alpha$ and [SII]/H$\alpha$ ratios, with CGCG097-120 consistently showing the highest value for these ratios. We explicitly verified the impact of an uncertain GANDALF stellar continuum subtraction on these results by measuring the line ratios in the extreme assumption of not applying any continuum subtraction and we found the line ratios to vary by less than 10\%. 

To better understand if the different line ratios arise due to on evolution of the RPS process, we focus next on whether the relative extension of the stellar and gas disks can be linked to these different line ratios.
Indeed, the degree of truncation of the disk has often been regarded as a proxy for the progress of the RPS event with more truncated disks being at a more advanced stage in the stripping \citep{Fossati13, Cortese12,Tonnesen12}.
Qualitatively, it is possible to see that the gas disk of CGCG097-079, CGCG097-073, and CGCG097-114 have a similar extent to the stellar disk. Conversely, in CGCG097-120 and in ESO137-001 the region characterized by ionised gas emission is between half to one third the size of the stellar disk. 
We quantify this by computing the $R_{{\rm H}\alpha}/R_{r{\rm -band}}$ ratio, where $R_{r{\rm -band}}$ is the radius at the 25$^{\rm th}$ mag arcsec$^{-2}$ of the stellar optical disk as previously seen in Fig.~\ref{fig:ratiomap}, and $R_{{\rm H}\alpha}$ is obtained from the H$\alpha$ emission line at a surface brightness limit of $\approx$ $2\times 10^{-18}$ erg s$^{-1}$ cm$^{-2}$ arcsec$^{-2}$, for all the objects. We now attempt to bring together the line ratio diagnostics and the $R_{{\rm H}\alpha}/R_{r{\rm -band}}$ ratio to identify if an evolutionary sequence can be identified in our sample.
\begin{figure}
	\includegraphics[width=\columnwidth]{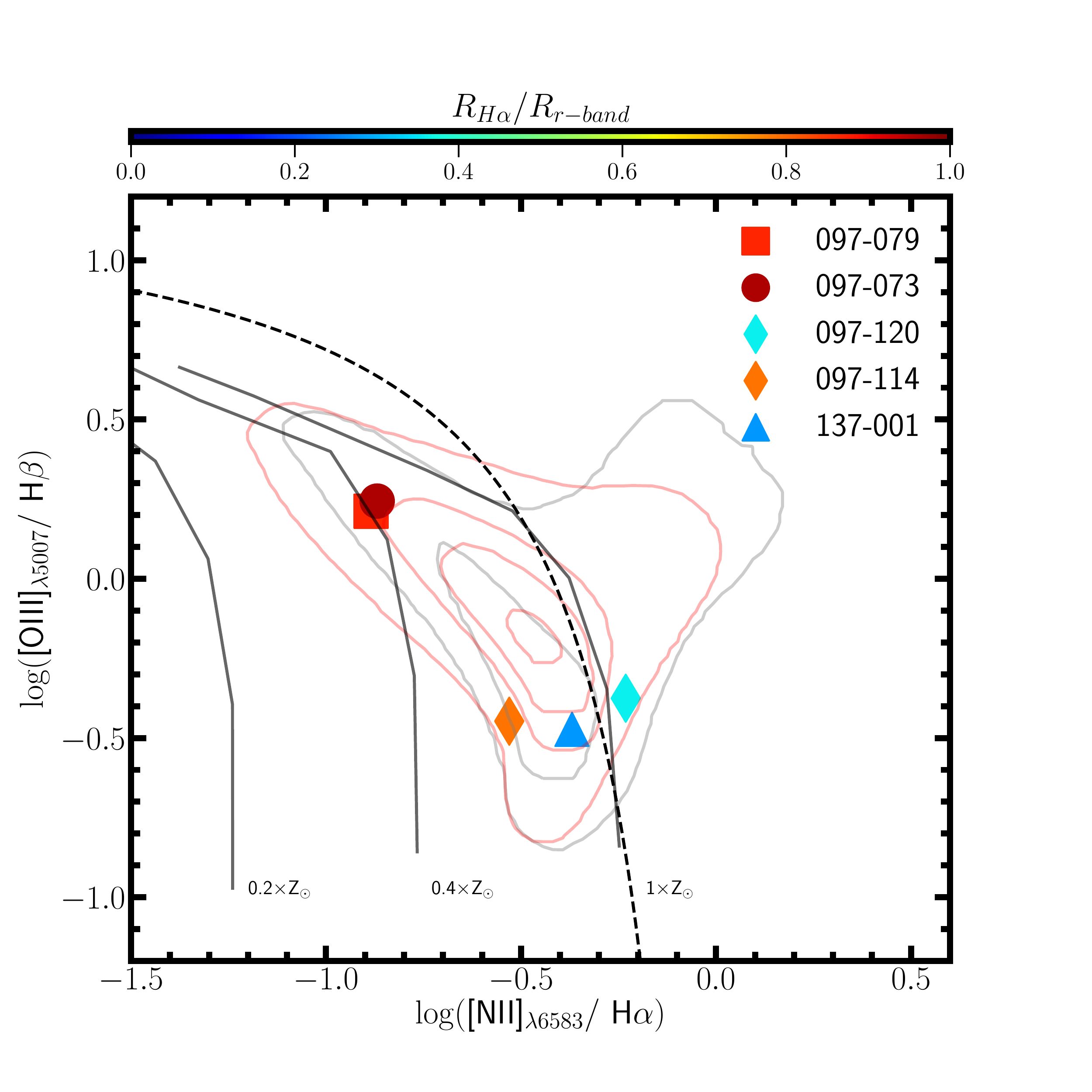}
    \caption{BPT diagnostic diagram [OIII]/H$\beta$ vs. [NII]/H$\alpha$ derived from the composite spectra extracted from the same regions of Fig.~\ref{fig:1Dbptgal}. Error bars are obtained from the KUBEVIZ fit. The colorbar represent the $R_{\rm H\alpha}/R_{r{\rm -band}}$ ratio. The grey and pink contours, and the solid and dashed black lines are as in Fig.~\ref{fig:bpt}.}
    \label{fig:trunqratio}
\end{figure}

In Fig.~\ref{fig:trunqratio} we show the BPT diagram of the [OIII]/H$\beta$ vs [NII]/H$\alpha$, as already plotted in the left panel of Fig.~\ref{fig:1Dbptgal} and which appears to maximise the differences in the emission line properties of the five galaxies in our sample. We color code the points extracted from the five interface regions of the galaxies by their value of $R_{{\rm H}\alpha}/R_{r{\rm -band}}$. It is immediately clear that the points in the top left region of the diagram show no or negligible truncation of the gaseous disk (i.e. values close to unity for the $R_{\rm H\alpha}/R_{r{\rm -band}}$ ratio). Conversely, CGCG097-120 and ESO137-001 show a significant truncation of the disk, with values of $R_{\rm H\alpha}/R_{r{\rm -band}}\approx 0.36$ and $\approx 0.27$, respectively. 
We thus conclude that the two galaxies with high [OIII]/$\rm H\beta$ appear to be undergoing a starburst at an early phase of the stripping process when they are still rich in cold/warm gas, while CGCG097-120 and ESO137-001 are already at a later stripping stage -- as shown by the significant truncation -- and their activity in the interface region is characterized by low levels of star formation and a more prominent contribution from shocks. Moreover, as shown in Table \ref{tab:sample}, CGCG097-120 exhibits the highest HI-deficiency of our sample indicating a significant lack of cold HI gas in its disk. An intermediate phase is shown by CGCG097-114, which does not show a starburst while still having an extended gaseous disk and a rich HI-content. We will discuss in detail the implications of these results in the following section.

\section{Discussion}
\label{sec:discussion}

Thanks to the increasing number of MUSE observations of active galaxies in the past years, we are now in the position of building a statistical sample of galaxies with the purpose of reaching a more complete view of the RPS events in the active cluster A1367. The rich multi-wavelength dataset available and the presence of an high fraction of blue galaxies infalling in the cluster at present days, make A1367 an ideal environment to better understand how RPS can actively remove the ISM of galaxies and what are the typical timescales leading to the quenching of the star formation activity. 

With our eight galaxies, we can identify similarities and differences both for the properties inside the galaxies and their ionised gas tails, also taking into account  morphology, location in the cluster, and line of sight velocity, with the ultimate goal of understanding the evolutionary stages of RPS. 
We now discuss the main effects of RPS on the activity of the galaxies and the physical conditions of their ionised gas tails.

\subsection{Ram-pressure stripping in CGCG097-79 and CGCG097-073}  
\label{sec:dis7379}
By focusing on the two galaxies CGCG097-079 and CGCG097-073, which we presented in this paper, it is possible to notice that both galaxies show ordered rotation with a velocity gradient (particularly obvious in CGCG097-079) in the direction of the gas flow. Furthermore, from the study of the velocity dispersion we can assert that there are no significant regions of high turbulence ($\sigma>75 \mathrm{ km~s^{-1}}$) in the disk and in the tail of the two galaxies. 
\citet{Gavazzi01} and \citet{Boselli&Gavazzi14} argued that the morphology at the far end of the tail of CGCG097-079, which becomes less streamlined and more clumpy, is suggestive of an interaction between the galaxies. With our MUSE data, we do not detect significant variations in the kinematics of the ionised gas in this region. While this does not rule out conclusively an interaction, the lack of kinematic signature does not offer support for this hypothesis, favouring instead a superposition in projection of these galaxies.

The diffuse gas filaments in these two galaxies do not exhibit bright and compact ionised gas regions, which are identified as HII regions in other RPS galaxies. While two regions can be identified in the tail of CGCG097-073 using the criteria detailed in \citet{Fossati16}, these are relatively extended ($FWHM \approx 2-3$ times the MUSE PSF) and are a factor of 10 fainter than the HII regions found in ESO137-001 \citep{Fossati16} or in CGCG097-087 \citep{Consolandi17}. These properties make them only candidate HII regions. Conversely, no source is identified in the tail of CGCG097-079. We will further discuss this piece of evidence in Section \ref{sec:disc_sf_tails}. The line ratios in these tails is typical of galaxies subject to RPS and confirms that they are undergoing a significant stripping process in the cluster hot halo. On the other hand, the compression of the gas in the leading edge of the two galaxies leads to a significant burst of star formation (Fig.~\ref{fig:ratiomap}), which results in the global SFR of the two galaxies being roughly a factor of 10 above the main sequence of galaxies of the same mass.

\subsection{Star formation in the disks of RPS galaxies}

Several studies have shown that galaxies in clusters lose their gas via RPS. The lack of atomic gas and the decrease of the molecular component induce a decrease of the star formation activity \citep{Balogh04, Baldry06, Weinmann06, Boselli&Gavazzi06, Boselli&Gavazzi14, Fumagalli09, Gavazzi10, Wilman10}. However, observations and simulations suggest that under certain conditions the SFR can be enhanced, generating a starburst \citep{Donas90, Fossati12, Vulcani18, Poggianti19, Boselli21}. Moreover, observations  \citep{Gavazzi95,Vollmer03, Boselli21} and simulations \citep{TroncosoIribarren16} show SF enhancement only in the interface region of the galaxies, suggesting that RPS is responsible for a compression at the leading edge of the stellar disk and a temporary increase in the SFR.
In this scenario, it is then possible to build an evolutionary sequence, in which galaxies infalling in clusters experience first a localized boost in the SFR, due to gas compression. After this enhancement, galaxies continue to lose their gas via RPS, finally quenching their star formation activity.

The different features observed in our sample of galaxies suggest that ionised gas in RPS interface regions shows different star formation properties at different stages of evolution. 
The increasing in [OIII]/H$\beta$ ratio observed in CGCG097-079 and CGCG097-073 could be a result of the SFR boost caused by compression of the ionised gas due to ram-pressure stripping, as also indicated by the presence of several bright star forming knots in arc-shaped regions as shown in Fig.~\ref{fig:73} and Fig.~\ref{fig:79} \citep[see also][]{Gavazzi95}.
On the other hand, high values of [NII]/H$\alpha$ and low [OIII]/H$\beta$ ratio in CGCG097-120, CGCG097-114 and ESO137-001 may be caused by a more advanced stripping phase when the outer regions of the galaxies have been completely emptied of gas and the star formation process has almost stopped. In this case shocks could be dominating the residual ionization of the gas.
As it can be seen in Fig.~\ref{fig:ratiomap}, the inner parts of these disks (particularly CGCG097-120) show line ratios consistent with photo-ionization which is indicative that the stripping process has not yet reached those inner regions. Furthermore, the progress of the RPS event can be seen from the amount of truncation of the star forming disk \citep{Fossati13}. For CGCG097-120 and ESO137-001 (see Fig.~\ref{fig:trunqratio}) the stellar disk is approximately twice as large as the star forming region. A similar feature is visible in NGC4330 in the Virgo cluster \citep{Fossati18} where ram pressure has removed the outer gas in the past 500 Myr and is now proceeding in the inner half of the stellar disk extension. Galaxies presenting low truncation in the disk, such as CGCG097-073 and CGCG097-079, settle in the top left region of the BPT diagram where they experience an enhancement in the SFR. A disk truncation is not observed in CGCG097-114, even if it shows shock-like emission-line properties, making this object possibly in an intermediate stage between the gas compression phase and the truncation and quenching phase.

\begin{figure}
	\includegraphics[width=\columnwidth]{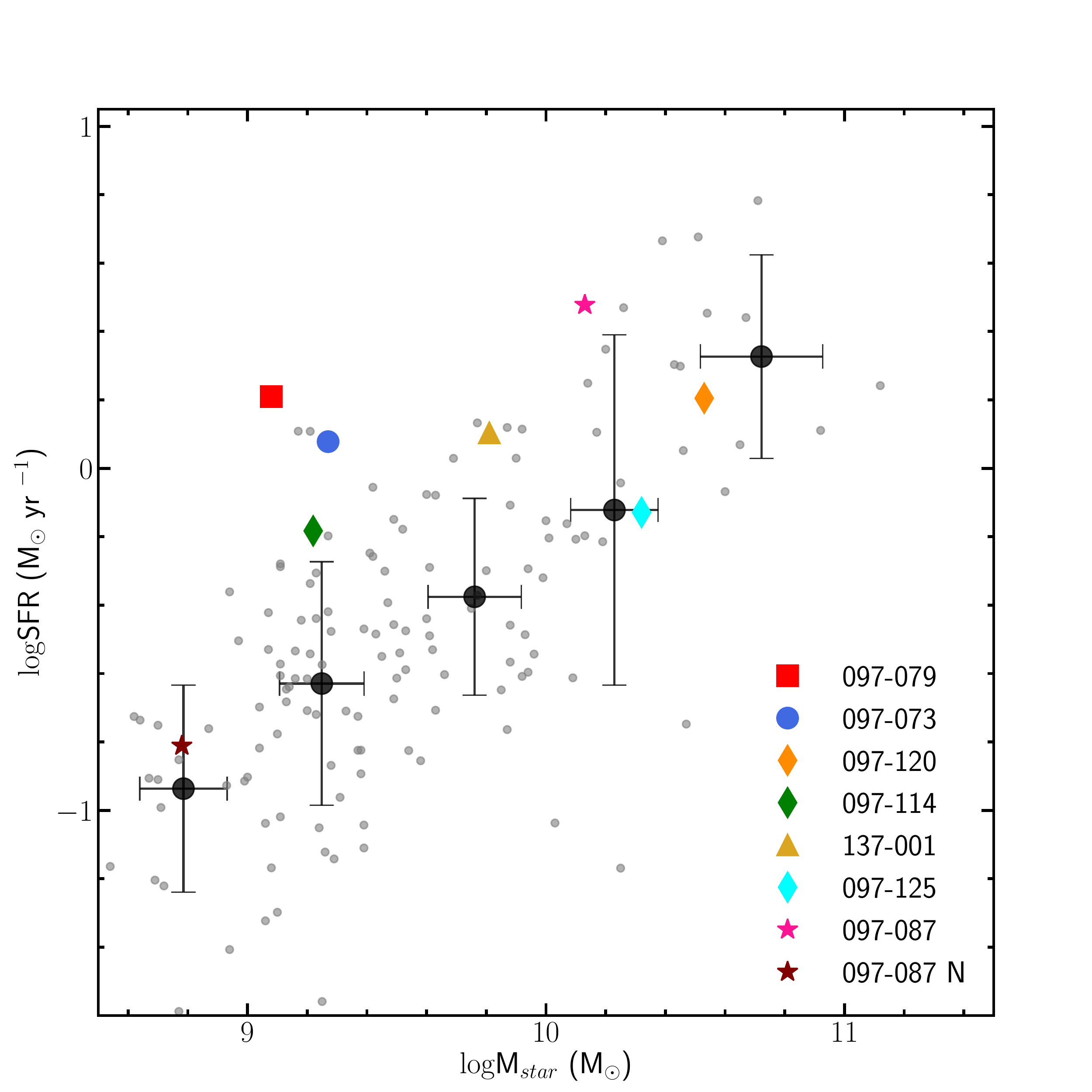}
    \caption{Relationship between the star formation rate and the stellar mass (main sequence) for our sample of galaxies suffering RPS. The grey filled dots represent the mean main sequence relation extracted from the \textit{Herschel} Reference Survey (HRS) for HI-normal galaxies \citep{Boselli15} and the black filled circles give the mean values and the standard deviations in different bins of stellar mass for these points.}
    \label{fig:MS}
\end{figure}

In Fig.~\ref{fig:MS} we show the relationship between the star formation rate and the stellar mass (main sequence, MS) for our sample of RPS galaxies, compared with the main sequence relation from the \textit{Herschel} Reference Survey (grey filled dots) for HI normal galaxies which represent an unbiased sample of unperturbed galaxies at comparable redshifts \citep{Boselli15}. In both samples, SFR values are obtained using the dust corrected H$\alpha$ flux, limiting the systematic offsets that could arise when different tracers are used. Furthermore, as shown in Table \ref{tab:sample}, our RPS galaxies do not host a central AGN emission that could affect the SFR estimates, and in the case of CGCG097-120 which is classified as a LINER, we remove the central emission when computing the SFR.

CGCG097-079 and CGCG097-073 stand out as significant outliers, being $\approx 1 $ dex above the main sequence and are undergoing a strong starburst. CGCG097-087 is also above the main sequence but its edge-on morphology makes the analysis of the interface region and of the disk truncation more difficult. The other galaxies are generally on or slightly above the MS. This is not unexpected as even for the most truncated galaxy in our sample (CGCG097-120) the amount of star formation taking place in the outer disk is small compared to the inner disk. Other galaxies in the Virgo cluster, for instance NGC4569 \citep{Boselli16B} and NGC4330 \citep{Fossati18}, show more truncated disks and a global SF activity that is a factor of 5-10 below the MS. We therefore expect that the galaxies in our A1367 sample, as well as the aforementioned Virgo galaxies, will become passive objects in the next $1-2$ Gyr \citep{Fossati18}. Moreover, the fact that only a fraction of galaxies show an active starburst could be an indication that this is a relatively short-lived phase at the beginning of the stripping process and it is followed by a decrease in the star formation activity leading to quiescence. This is consistent with the study of a complete sample of galaxies in the Virgo cluster by \citet{Boselli16}. These authors found that RPS could be responsible for the majority of the transformations of star-forming objects into passive ones in the cluster environment.

\subsection{Properties of the RPS tails}
\label{sec:tails}

The H$\alpha$ emission line is mostly used to detect extended ionised gas around galaxies at a typical gas temperature of T $\simeq$ 10$^4$ K \citep{Osterbrock06}. All the objects we followed-up with MUSE exhibit extended ionised gas tail, with the only exception of CGCG097-120 where the RPS event is seen almost face-on \citep{Fossati19}.

Due to the extension of the tails, we have been able to study the ionization conditions of the gas, using the BPT diagrams (Fig.~\ref{fig:1Dbpt}), in 14 independent spatial regions. A clear evidence stands out: 13/14 regions have an elevated [OI]/H$\alpha$ ratio combined with a low [OIII]/H$\beta$ ratio, placing them on the right of the photoionization separation line. These ratios are highly unusual both for the nuclei and the outer disks of star-forming galaxies occurring in $<2\%$ of the datasets drawn from these surveys. In our sample instead, the fraction of tails with elevated [OI]/H$\alpha$ is close to 100\%. We therefore argue that these ratios are a very typical signature of the diffuse gas emission of RPS events.

The observed line ratio cannot be explained by photoionization alone or by classic shock-heating models, as first discussed in \citet{Yoshida12}. These authors conclude that the low-velocity shocks are the most likely origin of this line diagnostic signature. We cannot however rule out the contribution of heat conduction or radiation from the hot ICM and of magneto-hydrodynamic waves or cosmic-rays in generating this distinctive signature. Conversely, the tail regions cluster in the photoionization region of the [SII]/H$\alpha$ diagram making the [SII] line less sensitive to the aforementioned processes than [OI]. Lastly, the individual data points are scattered over more than one order of magnitude in the [NII]/H$\alpha$ ratio, probably because this ratio is more sensitive to the gas metallicity \citep{Wuyts16}. Despite this scatter, all the points are in the photoionization part of the [NII]/H$\alpha$ BPT diagram. 
We note that the "Diffuse Ionized Gas (DIG)" associated with some nearby star forming galaxies shows similar emission-line ratios (see e.g. \citealt{Rand98}). \citet{Binette09} tried to explain the emission-line spectrum of the DIG in NGC891 with a combined model of turbulent mixing layer and  photoionization, reproducing the emission-line ratios except for [NII]/H$\alpha$ ratio (which could be modulated by metallicity). However, their model cannot reproduce the luminosity of the emission lines, and it remains unclear whether the same physical conditions apply in a completely different environment like the tails of RPS galaxies. Moreover, we can observe the same ionization features in the orphan cloud in A1367 that cannot be traced to a parent galaxy and has an unclear origin \citep{Ge21B}. \citet{Poggianti19} found similar trends for the [NII]/H$\alpha$ and [OI]/H$\alpha$ ratios in a sample of RPS tails behind 16 galaxies from the GASP survey. The authors conclude that the ionization mechanism is mostly photoionization. Our results make clear that diagnostic diagrams have to be interpreted with caution and that while some lines show ratios consistent with photoionization, the role of shocks or other processes cannot be neglected due to the peculiar [OI]/H$\alpha$ ratios founds in these tails. We also note that the GASP sample include galaxies selected to have star formation activity in the tails, as a result, the dominance of photoionization could be higher than in our sample of galaxies selected solely on the presence of H$\alpha$ tails. 

Focusing on the kinematics, our sample of tails mostly show ordered rotation consistent with the gas kinematics when it was still located in the galaxy disk \citep{Fumagalli14}. The velocity dispersion is mostly below 50 $\rm{km~s^{-1}}$ and it increases only at large distances from the galaxies where turbulence starts to dominate over the ordered motion. Similar features are also observed in the MUSE data of galaxies subject to RPS in the Virgo cluster \citep{Boselli21} or in galaxies in the GASP survey \citep{Bellhouse17}. The ordered motion of the gas in the tail could be an additional evidence of the RPS mechanism and can be used to disentangle pure RPS events from other cases where gravitational interactions are at play (e.g. CGCG097-125 in BIG, \citealt{Fossati19}). \citet{Vulcani21} provides a complete checklist to identify RPS events from other environmental mechanisms. The presence of a companion is usually a clear indication of gravitational interactions, however in the case of CGCG097-125 a companion might not be obviously found as the galaxy appears to be a post-merger remnant. Similarly, high speed encounters in the cluster environments could lead to a companion being at large distances from the perturbed galaxy. In summary, the presence of a ionised gas tail, evidence of shocks and of ordered motion of the galaxy stellar disk and of the tail are excellent indications of RPS. However, in the rich environment of clusters it is very likely that RPS galaxies are also perturbed by other mechanisms, further increasing the complexity of the interpretation.

\subsubsection{Star Formation in tails}
\label{sec:disc_sf_tails}

\begin{table}
\caption{The fraction of H$\alpha$ flux in compact knots (putative HII regions) to the total flux in the tails and filaments of the galaxies in our sample. The samples of compact knots are taken from \citet{Fossati16,Consolandi17,Fossati19} for the literature sample presented in these papers; and are extracted using the same procedures for the two galaxies first presented in this work: CGCG097-073 and CGCG097-079. In the case of CGCG097-079 we identify no compact source in the tail and we assume the same flux of sources as in CGCG097-073 in computing the fraction, which we consider to be an upper limit.}
{
\centering
\begin{tabular}{lr}
\hline

Name      	& $f_{\rm{H\alpha, HII}}/f_{\rm{H\alpha, tail}}$ \\  
\hline
\hline

ESO137-001	& $0.18$ \\
CGCG097-087	& $0.30$ \\
CGCG097-073	& $<0.02$	\\ 
CGCG097-079	& $0.03$	\\
BIG F1 & $0.03$    \\
BIG F2 & $0.24$     \\
BIG F3 & $0.03$     \\
\noalign{\smallskip}
\hline
\end{tabular}

\label{tab:hiifrac}}
\end{table}

The presence of SF in filaments around galaxies \citep{Smith10} is a highly debated topic, with contrasting results in literature. \citet{Vulcani18} and \citet{Poggianti19} show that there is a systematic enhancement in SF both in tails and galaxy disks at redshift $0.04 <z< 0.1$ within the GASP survey. As mentioned above, this enhancement is likely to increase the photoionization rate and therefore the energy input into the diffuse gas in the tails. This effect could be more pronounced in the GASP galaxies due to their selection method, where galaxies are followed up with MUSE if they have an asymmetric distribution of compact young knots outside the galaxy disks. Indeed, CGCG097-079 and CGCG097-073, despite their long tails, only show a few faint and relatively extended knots (which we classify as candidate HII regions) without a continuum counterpart in the Subaru $B-$band image. As a result those galaxies would not satisfy the GASP selection. Conversely, ESO137-001 shows 33 compact SF knots mostly located in the inner part of the tail where the gas velocity dispersion is lower and the tail is younger \citep{Fumagalli14,Fossati16}, and CGCG097-087 presents SF knots throughout the entire extent of the tail covered by our MUSE observations \citep{Consolandi17}. Table \ref{tab:hiifrac} presents the fraction of H$\alpha$ flux in compact knots to the total flux in the tails and filaments of the galaxies in our sample. In our sample the fraction ranges from 2\% to 30\% in the tail of CGCG097-087. At face value, this result is not in contrast with the conclusions of \citet{Poggianti19} that the GASP RPS tails are powered by SF, since we cannot exclude that the diffuse gas in the tail is at least partially kept ionized by HII regions. However, the presence of extended tails with little or no detectable HII regions and the almost ubiquitous elevated [OI]/H$\alpha$ ratio in our tails could suggest a minor role of SF as ionization mechanism.

Our results are reinforced by observations of RPS tails in the Virgo cluster. NGC4569 has the most extended RPS tail in the cluster \citep{Boselli16B} but without detectable regions of star formation. \citet{Boissier12} analysed deep UV images of seven galaxies in the Virgo cluster with HI tails. These authors did not detect star formation in the tails which they interpreted with a lower star formation efficiency in these environments. With the samples that have been collected so far, it remains unclear what drives the presence or absence of compact knots of star formation in the tails. We postulate that the local ICM density and the inclination of the galaxy with respect to the wind could be relevant parameters leading to these differences but we cannot yet draw firm conclusions with the data at hand.

\section{Conclusions}
\label{sec:conclusion}
RPS is an efficient process for the removal of ionised gas from galaxies in rich clusters like A1367. The morphology and the turbulence of this cluster makes it an ideal environment to study the physics of RPS with high spatial resolution due to its small distance to us.
We presented new MUSE observations of a large mosaic of 8 pointings covering CGCG097-079 and CGCG097-073, two galaxy members of A1367, and their respective diffuse ionised gas filaments arising from RPS. This analysis builds on previous work presented in a series of papers \citep{Fumagalli14, Fossati16, Consolandi17, Fossati19}, which includes the study of galaxies suffering RPS in A1367, plus ESO137-001 which is falling in the Norma cluster. Despite this sample not being fully complete, it is large and homogeneous enough to put constraints on the physical properties of this type of galaxies and on the ionization conditions of their tails.

Focusing on the new observations of MUSE of CGCG097-079 and CGCG097-073 for which we presented an in-depth analysis of the kinematics and the ionization conditions of the gas, we show that these galaxies do not present significant regions of high turbulence in the disk and in their tails (within the limits allowed by the MUSE spectral resolution). Moreover, we do not find trace of compact star formation regions in the gas diffuse filaments. However, we do see a clear enhancement of SF in the leading edge of the two galaxies, due to the compression of the gas. The study of the emission line ratios in the tails show they are typical of galaxies subject to RPS. 

Adding these new results to previous work from  this series of papers, we obtain a better picture of the properties of galaxies falling in A1367.

Our analysis has produced the following results:
\begin{enumerate}
    \item From the analysis of the emission line ratios in the RPS interface regions and the relationship between the star formation rate and the stellar mass we find that different ionization features are connected to different stages of the evolution. We find that this possible evolutionary sequence is strictly related to the ionization properties of the gas and the level of the gas truncation in the disk. We identify three main phases of the interaction. In an early phase galaxies could experience a burst of star formation placing them above the MS and before any truncation of the gas disk takes place. In an intermediate phase galaxies show a low ionization parameter in their star forming disk with little or no truncation of the gaseous disk (these objects are generally on the MS). In a later phase, galaxies have strongly truncated gaseous disks possibly characterized by shocks at the edges of the disk and are moving below the star-forming MS. 
     
    \item The study of the emission line ratios in the tails in our sample of galaxies reveal a consistently elevated [OI]/H$\alpha$ ratio combined with a low [OIII]/H$\beta$ ratio in all the regions of the tails. This is possibly the strongest spectroscopic signature of RPS in the extended gas. This feature cannot be explained by photoionization alone or by classical shock-heating models, and it only occurs in a small fraction of outer galaxy disks, while being very typical of hydrodynamic processes due to RPS. We conclude that in our representative sample of galaxies, shocks (or other complex ionization phenomena) play the most important role in setting the ionization conditions of these  diffuse gas filaments.
    \item Lastly, we find that only in $\approx 50\%$ of our objects (including RPS galaxies in Virgo) we observe compact HII regions in the tails. It is still unclear what drives the presence or absence of compact knots of star formation in the tails but it is becoming evident that this feature is not found in all the RPS events.
\end{enumerate}

Our integral field spectroscopic follow-up of RPS galaxies in A1367 revealed many similarities but also key differences in a homogeneous sample of RPS objects in a single cluster. Despite these efforts our sample remains relatively small and additional follow-up programmes and future observations targeting other rich clusters will continue to add constraints on the physical parameters of these type of galaxies, ultimately bringing us closer to understand how RPS shapes the formation of the red sequence in rich clusters.

\section*{Acknowledgements}
We thank D.Wilman for help in developing the {\sc Kubeviz} software. We thank the anonymous referee whose comments have improved the quality of this manuscript. This work is based on observations collected at the European Southern Observatory under ESO programmes 60.A-9349(A), 60.A-9100(G), 095.B-0023(A), 096.B-0019(A) and 098.B-0020(A) and, in part, on data collected at Subaru Telescope, which is operated by the National Astronomical Observatory of Japan. This project has received funding from the European Research Council (ERC) under the European Union's Horizon 2020 research and innovation programme (grant agreement no 757535) and by Fondazione Cariplo (grant No 2018-2329). MS acknowledges the support from the NSF grant 1714764.

\section*{Data Availability}

The raw and reduced data used in this work are available from the ESO Science Archive. Post-processed data and analysis codes are available from the corresponding author upon request. The {\sc Kubeviz} software is publicly available at: \url{https://github.com/matteofox/kubeviz}.



\bibliographystyle{mnras}








\bsp	
\label{lastpage}
\end{document}